\documentclass[aps,prb,twocolumn,superscriptaddress]{revtex4-1}

\usepackage{epsfig}
\usepackage{amssymb} 
\usepackage{amsfonts} 
\usepackage{mathtools} 
\usepackage{dcolumn} 
\usepackage{graphicx} 
\usepackage{color}  
\usepackage{bm}  
\usepackage{comment} 
\usepackage{units}
\usepackage{sidecap} 
\usepackage{textcomp} 
\usepackage[export]{adjustbox} 

\DeclarePairedDelimiter\abs{\lvert}{\rvert} 

\newcommand{\bpm}{\begin{pmatrix}}
\newcommand{\epm}{\end{pmatrix}}
\newcommand{\be}{\begin{eqnarray}}
\newcommand{\ee}{\end{eqnarray}}
\newcommand{\ba}{\begin{array}}
\newcommand{\ea}{\end{array}}

\def \ybco{YBa$_2$Cu$_3$O$_{6+\delta}$}
\def \hbco{HgBa$_2$CuO$_{4+\delta}$}

\begin{document}
\title{Angle-dependent magnetoresistance as a probe of Fermi surface warping in \hbco}

\author{Sylvia K. Lewin}
\affiliation{Department of Physics, University of California, Berkeley, California 94720, USA}
\affiliation{Materials Sciences Division, Lawrence Berkeley National Laboratory, Berkeley, California 94720, USA}

\author{James G. Analytis}
\affiliation{Department of Physics, University of California, Berkeley, California 94720, USA}
\affiliation{Materials Sciences Division, Lawrence Berkeley National Laboratory, Berkeley, California 94720, USA}

\date{\today}

\begin{abstract}
We develop a model for the angle-dependent magnetoresistance of \hbco\,in the underdoped regime where the Fermi surface is thought to be reconstructed by an ordered state such as a charge density wave. We show that such measurements can be employed to unambiguously distinguish the form of the Fermi surface's interlayer warping, placing severe contraints on the symmetry and nature of the reconstructing order. We describe experimentally accessible conditions in which our calculations can be put to the test.

\end{abstract}

\maketitle

\section{Introduction}

Characterizing the Fermi surface of the underdoped cuprates is a major challenge in understanding their strange metal properties. Quantum oscillations in \ybco\ (YBCO) first revealed that these materials have a coherent and measurable Fermi surface at magnetic fields above H$_{c2}$~\cite{Doiron-Leyraud2007}. Since this initial study, quantum oscillations have now also been observed in underdoped YBa$_2$Cu$_4$O$_{8}$ and \hbco\ (Hg1201); these measurements all indicated the presence of small Fermi surface pocket(s), presumably arising from Fermi surface reconstruction due to electronic order~\cite{Doiron-Leyraud2007, Bangura_YBCO_2008, Yelland_YBCO_2008, Barisic2013,Chan2016}. Charge density wave (CDW) correlations have been directly observed in both YBCO and Hg1201 through x-ray scattering~\cite{Ghiringhelli2012a,Chang2012,Tabis2014}, making this an attractive candidate for the origin of the reconstruction.  However, the details of the symmetry and shape of these small pockets, which is a major determining factor of normal state properties, remains unknown. In this study, we propose that angle-dependent magnetoresistance (ADMR) can reveal clear signatures of the morphology of the Fermi surface, and we propose straightforward experiments that can be performed in available magnetic fields.

\begin{figure}
	\centering
	\begin{tabular}[b]{l}
	\large (a)
	\includegraphics[trim=1.6cm 1.3cm 1.8cm 1.2cm, clip,width=0.36\linewidth,valign=t]{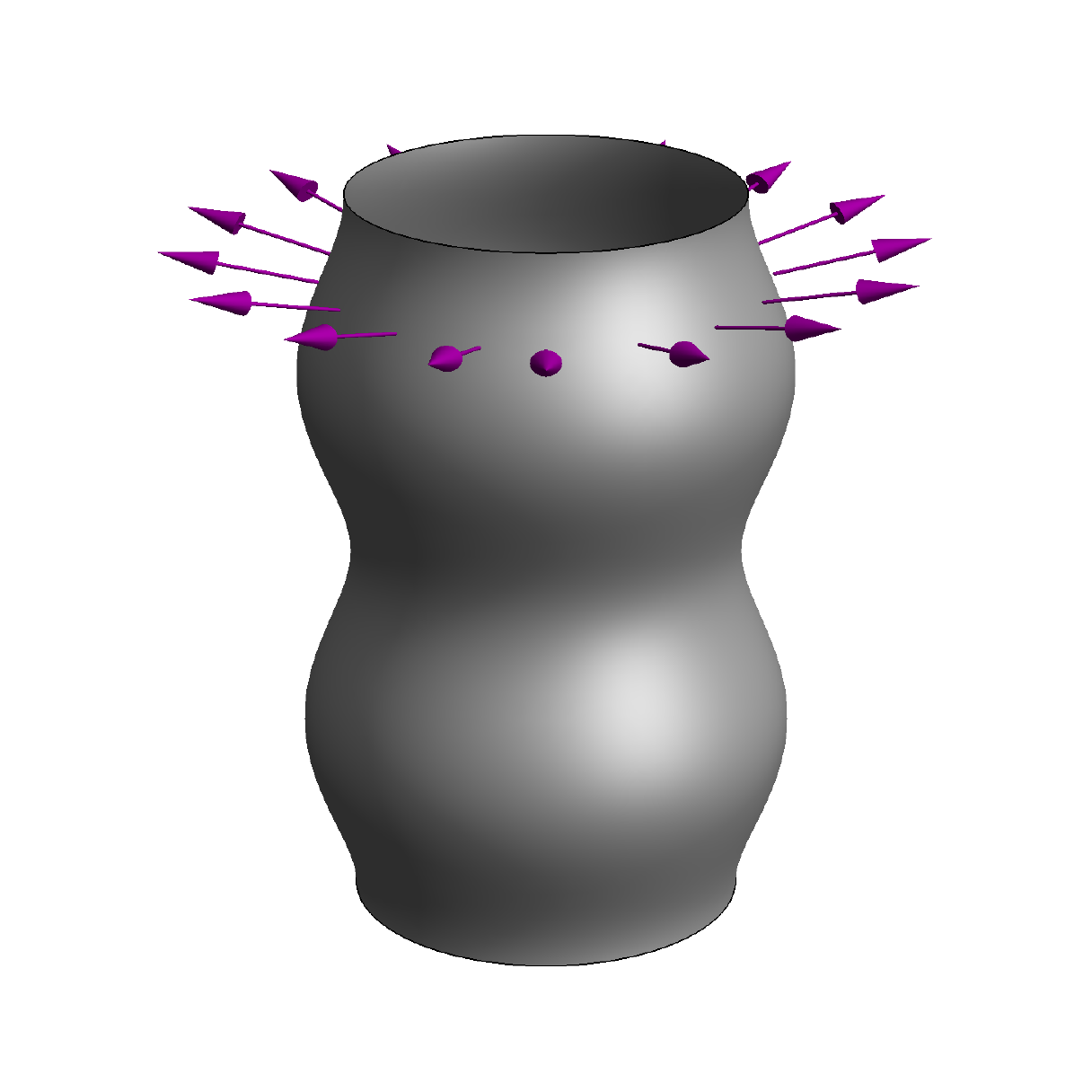}
	\end{tabular} \qquad
	\begin{tabular}[b]{l}
	\large (b)
	\includegraphics[trim=1.6cm 1.3cm 1.8cm 1.2cm, clip,width=0.36\linewidth,valign=t]{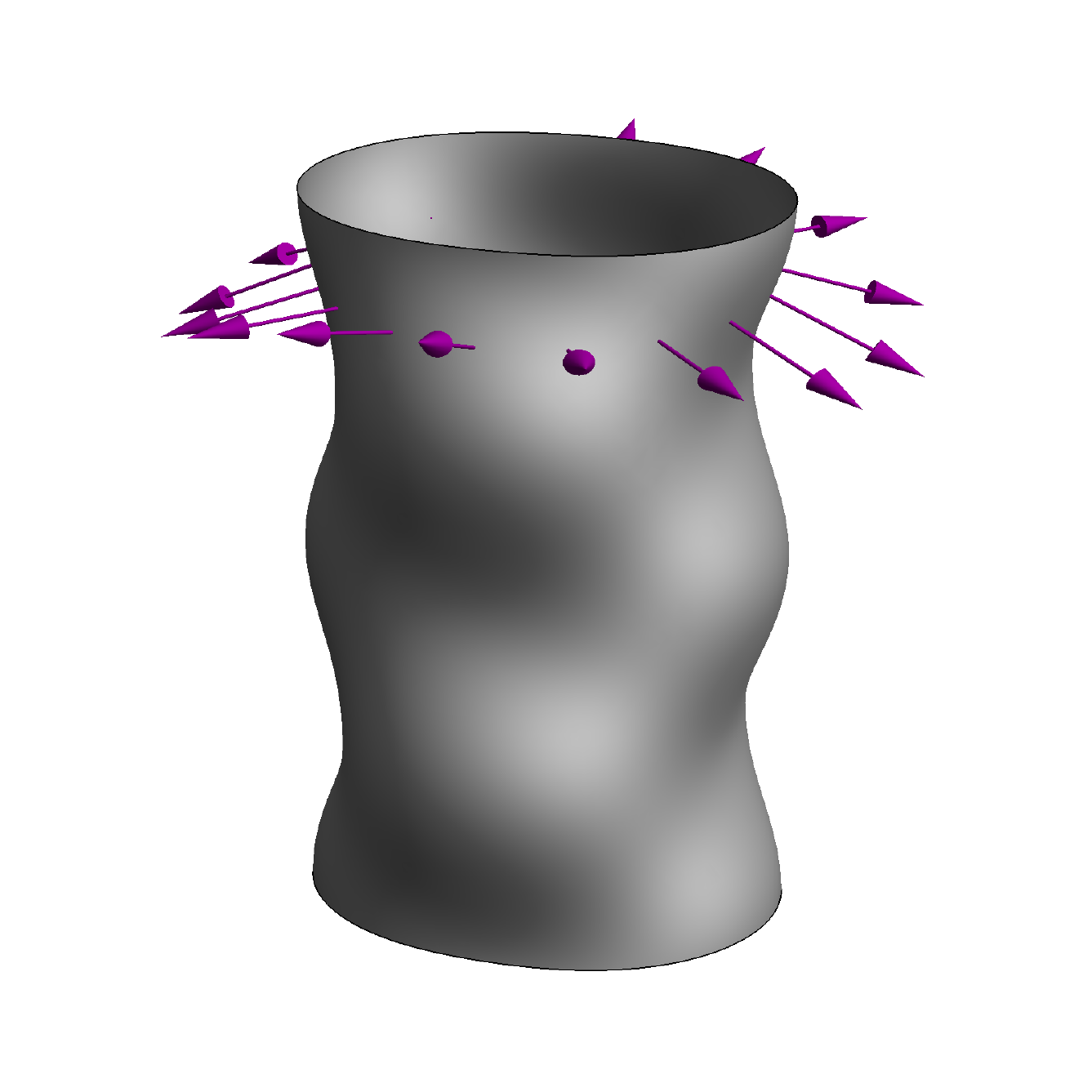}
	\end{tabular} \qquad
	\caption{(Color online) A quasi-two-dimensional Fermi surface with a) simple cosine interlayer warping and b) staggered twofold interlayer warping.  In both cases, the in-plane Fermi surface is circular before interlayer warping is imposed, and two periods of the interlayer warping are shown for clarity.  The arrows extending from each figure indicate the direction of the Fermi velocity for those points on the Fermi surface.}
	\label{fig:interlayerwarpings}
	\end{figure}

The layered structure of the cuprate superconductors leads to a quasi-two-dimensional Fermi surface in the form of a warped cylinder.  For a full understanding of these materials' properties and the order that drives their reconstruction, it is necessary to characterize not only the cross-sectional shape of the Fermi surface but also its interlayer warping. Sebastian \textit{et al.} performed detailed studies of Shubnikov-de Haas oscillations in YBa$_2$Cu$_3$O$_{6.67}$ and found that their data were consistent with a staggered twofold warping along $k_z$ (Fig. \ref{fig:interlayerwarpings}B) rather than a simple cosine warping (Fig. \ref{fig:interlayerwarpings}A)~\cite{Sebastian2014}. Their analysis suggested this could be caused by CDW order with $q\sim$($h,k,\nicefrac{1}{2}$), but recent scattering studies suggest that the ordering wavevector for the three-dimensional CDW is $q\sim$($h,k,1$) ~\cite{Gerber2015}. Furthermore, the quantum oscillation data can alternatively be explained without a twofold warping, and even without a three-dimensional Fermi surface~\cite{Maharaj2016,Briffa2016}.  This is thought to be possible because the mirror symmetry between the bilayer planes of YBCO is broken by CDW order~\cite{Forgan2015}.  The ambiguity in interpreting the quantum oscillation results highlights the need for a new symmetry-sensitive probe of the three-dimensional Fermi surface.

ADMR is a rich source of information about a material's Fermi surface geometry.  Under the influence of the Lorentz force, quasiparticles will trace out orbits on the Fermi surface that are perpendicular to the applied magnetic field.  As the angle of the applied field is varied, the quasiparticles sample different values of the Fermi velocity that vary in relation with the Fermi surface morphology, changing the observed resistance. Unlike quantum oscillations, ADMR is measured at a fixed field strength, meaning that it may even be used to probe field-induced changes in the Fermi surface. Moreover, for quasi-two-dimensional systems such as the cuprates, ADMR is calculated in the same manner regardless of whether the system has a coherent three-dimensional Fermi surface or incoherent interlayer transport (coherence is defined here as the requirement that the interlayer transfer integral $t_\perp$ is much greater than the scattering rate $\hbar/\tau$) ~\cite{Kennett2007}.\footnote{Coherent and weakly incoherent interlayer transport will have different signatures in ADMR only when the applied field is nearly in-plane, which is a regime that we have not included in these calculations\cite{Kennett2007}}   Therefore, ADMR can yield useful symmetry information in either case; the only difference is the interpretation of $t_\perp$.  For the remainder of this paper we will refer to the geometry of interlayer Fermi surface warping, but it is to be understood that we could alternatively be referring to the variation of interlayer hopping about the Fermi surface.

In this work, we will show that interlayer ($\rho_{zz}$) ADMR measurements of Hg1201 can be a powerful tool for characterizing the interlayer warping as well as the symmetry of the reconstructed Fermi surface.  Hg1201 has the highest transition temperature of the monolayer cuprates, has a simple tetragonal structure, and has relatively little disorder or buckling in the CuO$_2$ planes compared to other cuprate high-temperature superconductors~\cite{Barisic2008}.  Interlayer ADMR measurements of YBCO have been used to determine that its in-plane Fermi surface shape is most likely a square or diamond, which is consistent with a CDW reconstruction.\cite{Ramshaw2017}  However, the bilayer splitting in YBCO made it difficult to pinpoint the interlayer warping of the Fermi surfaces in that study.  Since Hg1201 only has a single copper-oxygen plane per unit cell, it avoids this complication. In addition, the copper-oxygen chains in the YBCO structure, which are known to affect transport properties at higher temperatures, are absent in Hg1201. All of these characteristics make Hg1201 an ideal model material for investigation with ADMR.

Using semi-classical calculations, we will show that a measurement of interlayer ADMR in Hg1201 at a single azimuthal angle can be used to clearly distinguish between the two types of interlayer warping shown in Fig. \ref{fig:interlayerwarpings}. From measurements at multiple azimuthal angles, even more symmetry information can be revealed. With strong enough magnetic fields, ADMR may also provide information about the periodicity of the interlayer warping and the shape of the in-plane Fermi surface. In Section \ref{sec:model} we describe the constraints placed on our model of Hg1201 from quantum oscillation experiments and we introduce the various Fermi surface geometries that we consider. In Section \ref{sec:calc} we outline the mathematical details of our model, deriving a general form for the interlayer conductivity of a quasi-two-dimensional metal under the constraints outlined in Section \ref{sec:model}. We calculate the expected interlayer ADMR in Section \ref{sec:results}, illustrating qualitative distinctions that arise from different Fermi surface geometries and particularly from different interlayer warpings. Finally, in Section \ref{sec:disc} we discuss the application of our ADMR results to actual measurements and define an experimentally accessible range of temperatures and magnetic fields in which our results can be applied.

\section{Experimental constraints and model parameters}
\label{sec:model}
Since Hg1201 is a layered, quasi-two-dimensional system, its Fermi surface must be in the form of a warped cylinder.  Quantum oscillation measurements in underdoped samples of Hg1201 have found a single frequency for each sample, in the range of 840-900 T for different dopings~\cite{Barisic2013,Chan2016}, which corresponds to roughly $3\%$ of the Brillouin zone area.  These measurements were taken with the magnetic field in the interlayer direction; that is, they measured the in-plane Fermi surface cross-section.  The results indicate that the Fermi surface consists of a single pocket (or multiple pockets of the same size).  The existence of a single quantum oscillation frequency also indicates that the interlayer warping cannot be large; in fact, an upper limit of $t_{\perp} < 0.35$ meV was determined, which is three orders of magnitude smaller than the intralayer nearest-neighbor hopping~\cite{Chan2016}.

\begin{figure}
    	\includegraphics[trim=1cm 1.4cm 0.7cm 0cm, clip, width=0.5\textwidth]{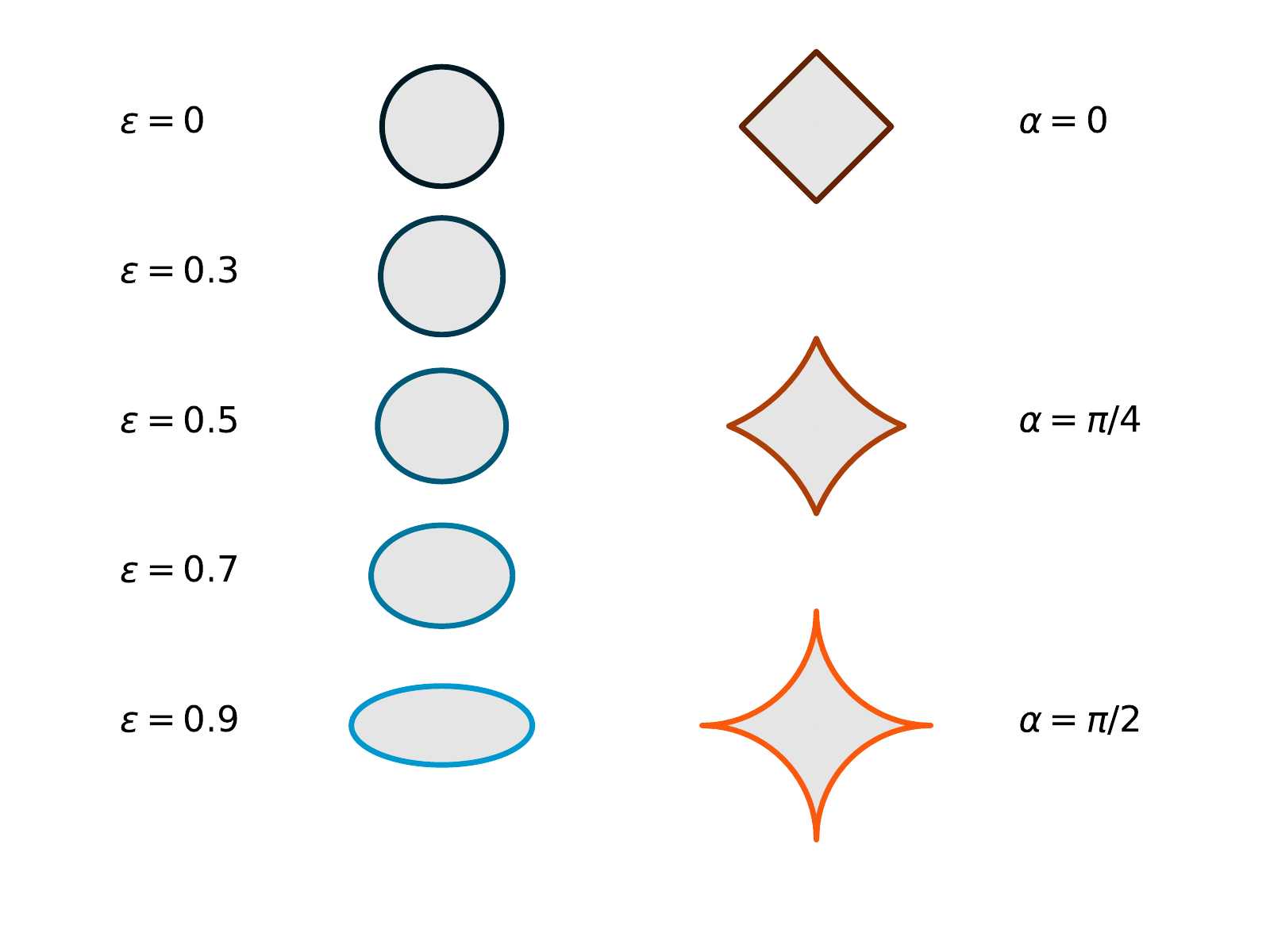}
  	\caption{(Color online) The various in-plane Fermi surface shapes considered in this paper.  The left column shows ellipses with increasing eccentricity ($\epsilon$) from top to bottom.  The right column shows diamond-shaped pockets with increasing concavity from top to bottom; $\alpha$ is the angle that must be subtended on a circle to create each arc that makes up the pocket.  Note that the area enclosed by all of these shapes is the same; we constrain the cross-sectional area to be consistent with quantum oscillation measurements, as described in the text.}
  	\label{fig:FSshapes}
	\end{figure}

The main result of our work is that the ADMR of Hg1201 can clearly distinguish cosine vs. staggered twofold interlayer warping.  We want to show that this is a qualitative distinction that can be made regardless of the details of the in-plane Fermi surface shape.  To that end, we produce simulations for a variety of in-plane Fermi surface shapes.  We consider a circular Fermi surface, the simplest option; various elliptical Fermi surfaces; and diamond-shaped pockets, ranging from square to concave circular arcs.  The choice to consider diamond-shaped pockets is motivated by the  hypothesis that the small Fermi surface pocket in Hg1201 is due to a CDW reconstruction, as described in Ref. \onlinecite{Chan2016}.  The full range of in-plane Fermi surface shapes we consider is illustrated in Fig. \ref{fig:FSshapes}.  As we vary the in-plane Fermi surface shape, we constrain its cross-sectional area such that it always corresponds to quantum oscillations of frequency 893 T, as reported for an underdoped sample with $T_c$ of 74 K~\cite{Chan2016}.

In addition to the Fermi surface geometry, ADMR is also affected by $\omega$, the angular frequency of a quasiparticle about the Fermi surface, and $\tau$, the quasiparticle scattering time.  For a perfectly cylindrical Fermi surface, the angular frequency is given by $\omega = \omega_0\cos(\theta)$, where $\theta$ is the angle of the applied magnetic field from the interlayer direction and $\omega_0$ is a constant.  Specifically, $\omega_0 = eB/m^*$, with $m^*$ being the quasiparticle effective mass (\textit{not} the cyclotron mass). The interlayer warping of our system is small, so it would still be appropriate to use this form of $\omega$ when considering a Fermi surface with a circular cross-section.  However, for the variety of in-plane Fermi surfaces we are considering, we must use a more general expression for $\omega$.  In a quasi-two-dimensional material with small interplane warping ($t_{\perp} \ll E_F$), we find
	\begin{equation}\label{eq:omegadef}
	\omega \approx \omega_0\cos(\theta)\cos\bm{(}\gamma(\varphi)\bm{)},
	\end{equation}
where $\gamma(\varphi)$ is the angle between the Fermi velocity and the Fermi momentum, which varies about the Fermi surface and is entirely determined by the Fermi surface geometry; see Appendix \ref{app:omegac} for the derivation of $\omega$.  For the purposes of our calculations, we do not care about $\omega_0$ or $\tau$ individually, but only $\omega_0\tau$.  For Hg1201, we expect $\omega_0\tau \approx 0.35$ at 45 T and at a temperature of a few degrees Kelvin~\cite{Chan2016}.  We perform calculations with $\omega_0\tau =$ 0.1, 1, and 10 to show how ADMR results can be expected to vary with changing magnetic fields and temperatures.

\section{Calculating interlayer ADMR}
\label{sec:calc}
\begin{figure}[h]
    	\includegraphics[trim=0.5cm 1.7cm 0.75cm 3cm, clip, width=0.34\textwidth]{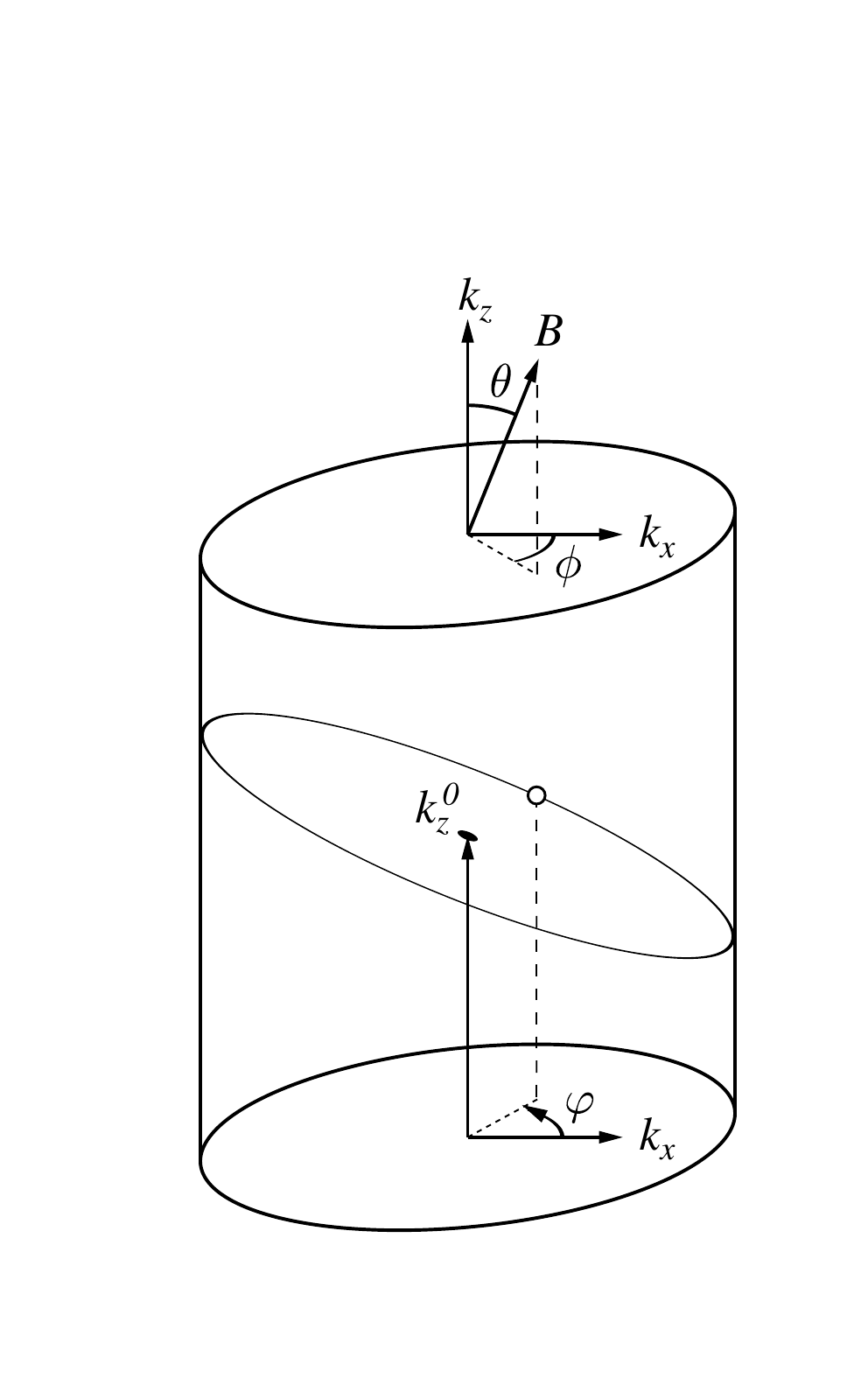}
  	\caption{The position of a quasiparticle, represented in this figure by an open circle, can be specified by the central height of the orbit it is tracing ($k_z^0$) and its azimuthal position ($\varphi$).  Note that $\varphi$ is defined with respect to $k_x$, as is $\phi$, the azimuthal angle of the applied magnetic field.  The polar angle of the applied field, $\theta$, is defined with respect to $k_z$.}
  	\label{fig:kz0}
	\end{figure}

We consider an applied magnetic field of strength $B$ whose direction is specified by a polar angle ($\theta$) with respect to $k_z$ and an azimuthal angle ($\phi$) with respect to $k_x$.  In the presence of such a field, quasiparticles in a quasi-two-dimensional material will trace out orbits on the cylindrical Fermi surface that are perpendicular to the field direction.  The quasiparticle position can therefore be specified by two coordinates: $\varphi$, which gives its azimuthal position, and $k_z^0$, which specifies the $k_z$ position of the center of the orbital plane; see Fig. \ref{fig:kz0}.

The conductivity of a quasi-two-dimensional material can be calculated semi-classically using the Shockley-Chambers tube integral form of the Boltzmann transport equation.  With a non-constant $\omega$ but a constant scattering rate $\tau$, this equation can be written as follows~\cite{Ziman1972, Abdel-Jawad2006}:
	\begin{equation}\label{eq:Shockley}
	\begin{split}
		\sigma_{\alpha\beta} = & \frac{e^3 B \cos(\theta)}{4\pi^3 \hbar^2} \int dk_z^0 \int_0^{2\pi} d\varphi_0 
		\frac{v_\beta(\varphi_0,k_z^0)}{\omega(\varphi_0)} \times \\
		&\int_{\varphi_0}^{\infty} d\varphi \frac{v_\alpha(\varphi,k_z^0)}{\omega(\varphi)}
			\exp\left(-\int_{\varphi_0}^{\varphi} \frac{d\varphi'}{\omega(\varphi')\tau}\right),
	\end{split}
	\end{equation}
where all velocities are taken to be at the Fermi surface.  See Appendix \ref{app:Shockley} for a derivation of this form of the equation.

The Fermi surface of a quasi-two-dimensional material can be described by the tight-binding form
\begin{equation}\label{eq:FS}
	E_F(k_z,\varphi) = \frac{\hbar^2 k_F^{2}(\varphi)}{2m^*} - 2t_{\perp}(\varphi)\cos(c k_z),
	\end{equation}
where \mbox{$k_F(\varphi)$} parameterizes the \emph{in-plane} Fermi surface, $t_{\perp}(\varphi)$ is the interlayer hopping integral, and $c$ is the interlayer lattice parameter.\footnote{This assumes that the Brillouin zone height is $\frac{2\pi}{c}$.  In, e.g., a body-centered tetragonal structure with Brillouin zone height $\frac{4\pi}{c}$ the cosine term would be $\cos\left(\frac{ck_z}{2}\right)$.}

Using $v_z = \frac{1}{\hbar}\frac{dE(\bm{k})}{d k_z},$ we find the interlayer velocity to be
	\begin{equation}
	v_z(\bm{k},\varphi) = \frac{2ct_{\perp}(\varphi)}{\hbar}\sin(ck_z).
	\end{equation}
Therefore,
	\begin{equation}\label{eq:Shockley1}
	\begin{split}
		\sigma_{zz} = & \frac{c^2 e^3B \cos(\theta)}{\pi^3 \hbar^4} \int dk_z^0 \int_0^{2\pi} d\varphi_0 
		\frac{t_{\perp}(\varphi_0)\sin(ck_z)}{\omega(\varphi_0)} \times \\*
		&\int_{\varphi_0}^{\infty} d\varphi \frac{t_\perp(\varphi)\sin(ck_z)}{\omega(\varphi)}
			\exp\left(-\int_{\varphi_0}^{\varphi} \frac{d\varphi'}{\omega(\varphi')\tau}\right). 
	\end{split}
	\end{equation}

The value $k_z$ should be more accurately written as $k_z(k_z^0,\varphi)$: a quasiparticle's position in $k$-space depends on the orbit it is following, as well as its azimuthal position on that orbit.  It turns out that the expression for $k_z$ also depends on whether the Fermi surface intersects the Brillouin zone.

If a Fermi surface is entirely contained within a Brillouin zone, each quasiparticle will stay on a single orbit through its lifetime and we can see from simple geometric considerations that
	\begin{equation}\label{eq:kzorbit}
	k_z = k_z^0 - k_F(\varphi)\tan(\theta)\cos(\varphi - \phi).
	\end{equation}

	\begin{figure}
    	\includegraphics[trim=0.5cm 1.2cm 0.75cm 0.7cm, clip, width=0.34\textwidth]{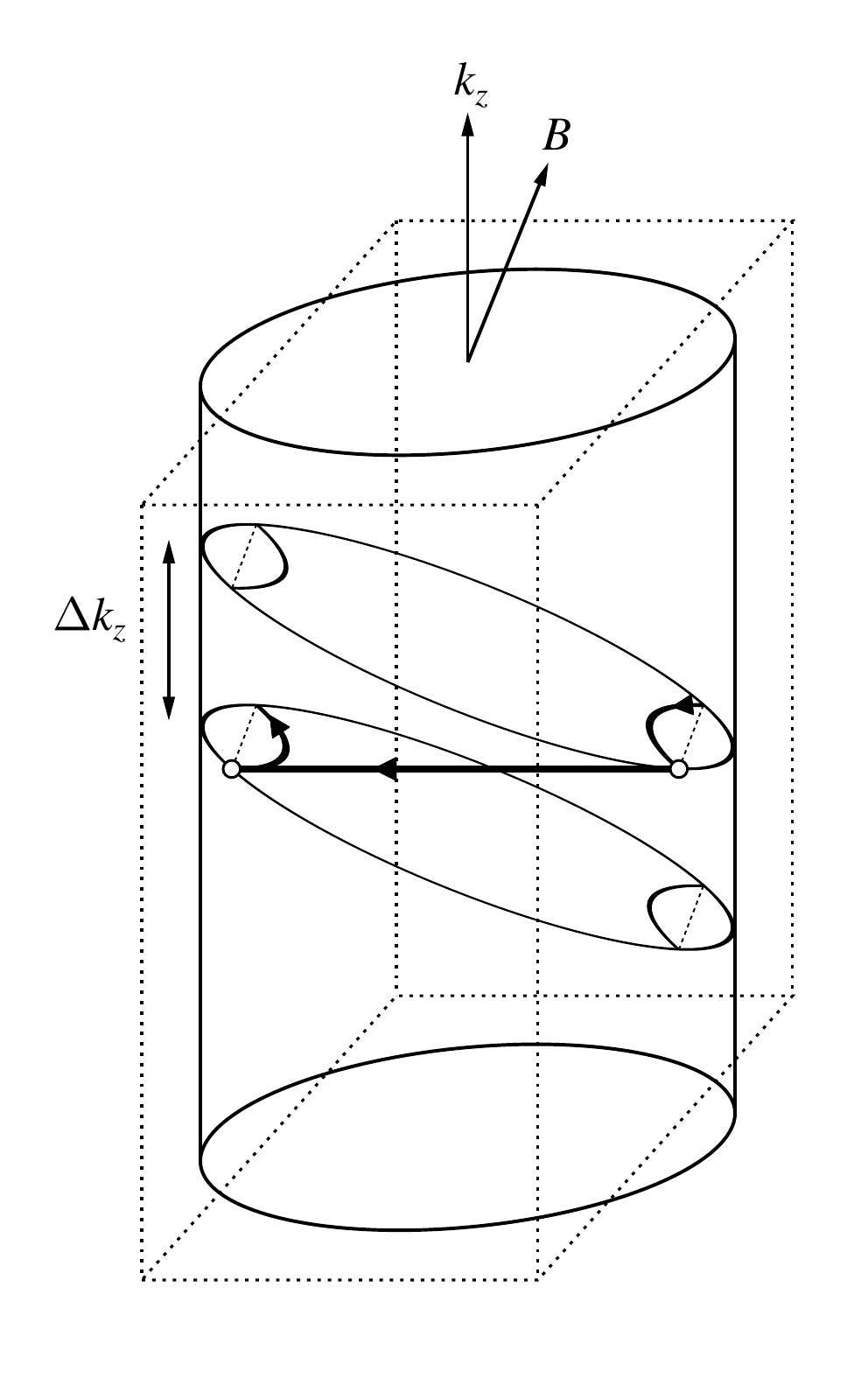}
  	\caption{When a quasiparticle reaches the edge of the Brillouin zone (indicated by dotted lines), it will typically undergo Bragg diffraction.  As illustrated above, the value of $k_z$ will be the same before and after Bragg diffraction but the value of $k_z^0$ will change by $\Delta k_z$ as the quasiparticle moves to a different orbit about the Fermi surface.  The value of $\Delta k_z$ depends on the geometry of the Fermi surface and the Brillouin zone, as well as the quasiparticle's azimuthal position on the Fermi surface when it undergoes Bragg diffraction.}
  	\label{fig:Bragg}
	\end{figure}

However, if the Fermi surface intersects the Brillouin zone boundary then quasiparticles will Bragg diffract at the zone boundary and will have a momentum change equal to a reciprocal lattice vector.  In this case, the value of $k_z$ will not change, but the value of $k_z^0$ will: the quasiparticle will jump to a different orbit about the Fermi surface, as illustrated in Fig. \ref{fig:Bragg}.  The amount by which $k_z^0$ changes will depend on the details of the Fermi surface and Brillouin zone geometries; it will also depend on the quasiparticle's azimuthal position when it undergoes Bragg diffraction.

As a further complication, each time a quasiparticle reaches the Brillouin zone boundary it has a non-zero probability, $p$, to avoid Bragg diffraction and effectively tunnel through the gap in $k$-space to stay on its original orbit~\cite{Cohen1961}.  This is a phenomenon known as magnetic breakdown; its probability in a quasi-two-dimensional material is given by $p = e^{-B_0/B\cos(\theta)}$, where $B_0$ is a material-specific parameter that is related to the size of the energy gap at the Brillouin zone boundary~\cite{Shoenberg}.   Because the quasiparticle can either Bragg diffract or undergo magnetic breakdown at each intersection of the Fermi surface and the Brilloun zone, these intersections are known as ``MB junctions."

Taking all of this into account, we can write
	\begin{equation}\label{eq:kzorbit_full}
	k_z = k_z^0 - k_F(\varphi)\tan(\theta)\cos(\varphi - \phi) + \sum\limits_j n_j(\varphi)\Delta k_z^{(j)},
	\end{equation}
where $j$ indexes all of the MB junctions, $n_j$ indicates the number of times the quasiparticle has Bragg diffracted at the $j^{\text{th}}$ MB junction, and $\Delta k_z^{(j)}$ is the amount by which $k_z^0$ changes when undergoing Bragg diffraction at the $j^{\text{th}}$ MB junction; $\Delta k_{z}^{(j)}$ can be found through simple geometric means, as described in Ref. \onlinecite{Lewin2015a}.    We set $n_j(\varphi_0) = 0$.  Note that in Eq. \ref{eq:kzorbit} and Eq. \ref{eq:kzorbit_full} we neglect the effects of interlayer warping, which should be small.

We can substitute this expression for $k_z$ into Eq. \ref{eq:Shockley1} and integrate over $k_z^0$ from $0$ to $\frac{2\pi}{c}$, the Brillouin zone height of Hg1201.  We then have

	\begin{equation}\label{eq:Shockley2}
	\begin{split}
		\sigma_{zz} = & \frac{c e^3B \cos(\theta)}{\pi^2 \hbar^4}  \int_0^{2\pi} d\varphi_0 
		\frac{t_{\perp}(\varphi_0)}{\omega(\varphi_0)} \int_{\varphi_0}^{\infty} d\varphi \frac{t_\perp(\varphi)}{\omega(\varphi)} \\*
	& \times \cos\bigg( - G(\varphi) +  c \sum\limits_j n_j(\varphi)\Delta k_z^{(j)} + G(\varphi_0)\bigg) \\*
	& \times \exp\left(-\int_{\varphi_0}^{\varphi} \frac{d\varphi'}{\omega(\varphi')\tau}\right), 
	\end{split}
	\end{equation}
where we have defined $G(\varphi) \equiv c k_F(\varphi)\tan(\theta)\cos(\varphi - \phi)$.

We can write $t_{\perp}(\varphi) = t_{\perp}^0 f(\varphi)$.  If we also insert Eq. \ref{eq:omegadef}, then for ease we can write our conductivity in units of $c e^3B (t_{\perp}^0)^2/\pi^2 \hbar^4 \omega_0^2$ so that it is dimensionless:

	\begin{equation}\label{eq:Shockley3}
	\begin{split}
		\sigma_{zz} = & \frac{1}{\cos(\theta)} \text{Re}\bigg[ \int_0^{2\pi} d\varphi_0 
		\frac{f(\varphi_0)}{\cos\bm{(}\gamma(\varphi_0)\bm{)}} \int_{\varphi_0}^{\infty} d\varphi \frac{f(\varphi)}{\cos\bm{(}\gamma(\varphi)\bm{)}} \\*
	& \times \exp\bigg(i[G(\varphi) - G(\varphi_0)] -ic\sum n_j(\varphi)\Delta k_z^{(j)} \\*
	& - \frac{1}{\omega_0\tau \cos(\theta)} \int_{\varphi_0}^{\varphi} \frac{d\varphi'}{\cos\bm{(}\gamma(\varphi')\bm{)}} \bigg) \bigg].
	\end{split} 
	\end{equation}

\begin{figure}
	\begin{tabular}[b]{l}
	\large (a)\\
	\includegraphics[trim=1.4cm 1.1cm 1.4cm 1.4cm, clip,width=0.44\linewidth,valign=t]{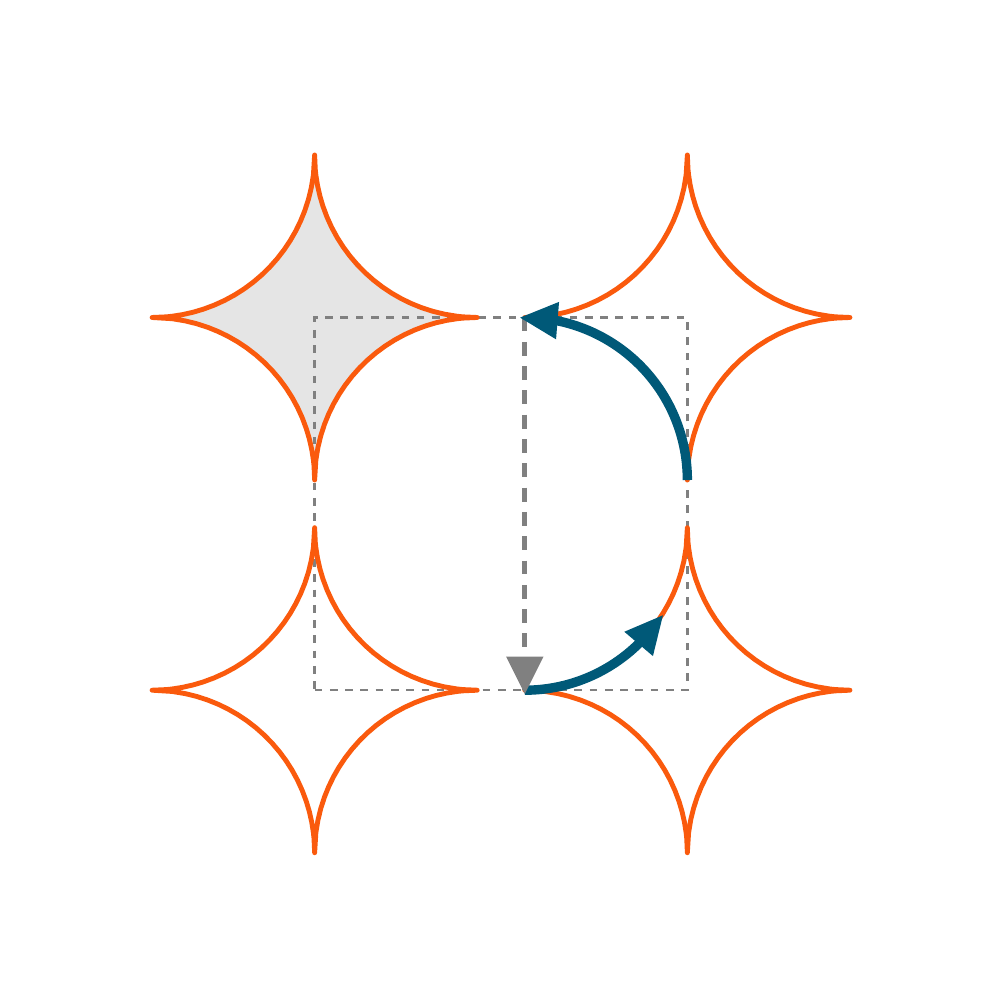}
	\end{tabular}\qquad
	\begin{tabular}[b]{l}
	\large (b)\\
	\includegraphics[trim=1.4cm 1.1cm 1.4cm 1.4cm, clip,width=0.44\linewidth,valign=t]{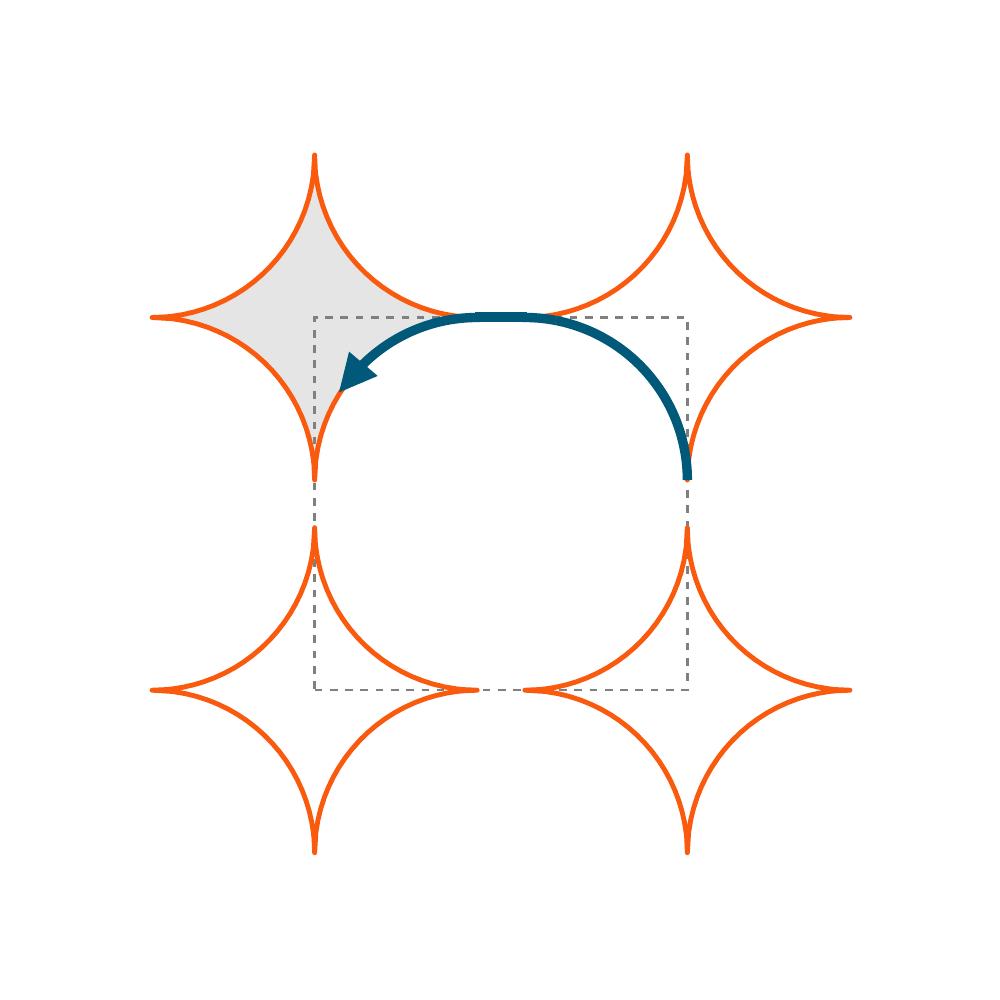}
	\end{tabular} \qquad
	\caption{(Color online) Concave diamond-shaped pockets at the corners of a Brillouin zone, as predicted for a biaxial CDW reconstruction.  The dotted line marks the Brillouin zone boundary.  Panel a) shows the path of a quasiparticle that undergoes Bragg diffraction, while panel b) shows the path of a quasiparticle that undergoes magnetic breakdown.  The shaded pocket in each panel is meant to emphasize the fact that even when considering magnetic breakdown effects, the area we fix to match quantum oscillation measurements is that of the diamond-shaped pockets.}
	\label{fig:BraggvsMB}
	\end{figure}

It is important to make use of the above equation, which allows for Bragg diffraction and magnetic breakdown, when considering the hypothesis that the small Fermi surface pockets in Hg1201 are the result of a biaxial CDW reconstruction~\cite{Chan2016,Tabis2014}.  If this is the case, there will be four diamond-shaped Fermi surface pockets in the corners of the reconstructed Brillouin zone~\cite{Harrison2014}, as illustrated in Fig. \ref{fig:BraggvsMB}.  Quasiparticles are expected to Bragg diffract when they reach a Brillouin zone boundary, as shown in Fig. \ref{fig:BraggvsMB}(a).  If Bragg diffraction always occurs, then the quasiparticles simply trace out the concave electron-like Fermi surface pockets.   But if a quasiparticle undergoes magnetic breakdown at the Brillouin zone boundary, it will instead traverse the larger hole-like Fermi surface, as shown in Fig. \ref{fig:BraggvsMB}(b).

In the simpler case of a Fermi surface that does not intersect the Brillouin zone boundary, we have
	\begin{equation}\label{eq:Shockley4}
	\begin{split}
		\sigma_{zz} = & \frac{1}{\cos(\theta)} \text{Re}\bigg[ \int_0^{2\pi} d\varphi_0 
		\frac{f(\varphi_0)}{\cos\bm{(}\gamma(\varphi_0)\bm{)}} \int_{\varphi_0}^{\infty} d\varphi \frac{f(\varphi)}{\cos\bm{(}\gamma(\varphi)\bm{)}} \\*
	& \times \exp\bigg(i[G(\varphi) - G(\varphi_0)] \\*
	& - \frac{1}{\omega_0\tau \cos(\theta)} \int_{\varphi_0}^{\varphi} \frac{d\varphi'}{\cos\bm{(}\gamma(\varphi')\bm{)}} \bigg) \bigg].
	\end{split}
	\end{equation}

The two situations require different methods of calculation, as described in the following sections.

\subsection{Fermi surface in a single Brillouin zone}

Taking advantage of the periodic nature of $f(\varphi)$, $\gamma(\varphi)$, and $G(\varphi)$, we can rewrite Eq. \ref{eq:Shockley4} as
	\begin{equation}\label{eq:Shockley5}
	\begin{split}
		\sigma_{zz} = & \frac{1}{1-P(0, 2\pi)} \frac{1}{\cos(\theta)} \text{Re}\bigg[ \int_0^{2\pi} d\varphi_0 
		\frac{f(\varphi_0)}{\cos\bm{(}\gamma(\varphi_0)\bm{)}} \\*
 &  \times \int_{\varphi_0}^{\varphi_0 + 2\pi} d\varphi \frac{f(\varphi)}{\cos\bm{(}\gamma(\varphi)\bm{)}}
 \exp\bigg(i[G(\varphi) - G(\varphi_0)] \bigg) P(\varphi_0, \varphi) \bigg],
	\end{split}
	\end{equation}
where 
\begin{equation}
P(\varphi_1, \varphi_2) \equiv \exp\left(- \frac{1}{\omega_0\tau \cos(\theta)} \int_{\varphi_1}^{\varphi_2} \frac{d\varphi'}{\cos\bm{(}\gamma(\varphi')\bm{)}}\right).
\end{equation}

The equation above can be integrated numerically; when possible, the integral in $P$ should first be solved analytically to make numerical integration more efficient.

\subsection{Fermi surface across the Brillouin zone boundary}

When the Fermi surface intersects the Brillouin zone boundary, the full version of Eq. \ref{eq:Shockley3} is needed.  As noted above, the term $n_j$ in this equation is a count of the number of times a quasiparticle has Bragg diffracted at an MB junction, where the different MB junctions are labeled by the index $j$.  Keeping track of the evolution of these $n_j$ is non-trivial, making it a challenge to evaluate Eq. \ref{eq:Shockley3}.  Building on Falicov and Sievert's treatment of magnetic breakdown for in-plane conductivity~\cite{Falicov1965}, Nowojewski \textit{et al.} showed that this issue can be solved by writing a self-consistent, vectorized form of the Shockley-Chambers tube integral~\cite{Nowojewski2008,Nowojewski2010}.  This method has previously been used to study systems wherein $\omega$ did not vary with $\varphi$~\cite{Blundell2010,Lewin2015a}.  For this work, we extend the equations of Ref. \onlinecite{Lewin2015a} to account for the variation of $\omega$ about the Fermi surface.  We write the (dimensionless) conductivity in the following form:

	\begin{equation}\label{eq:vectorShockley}
	\sigma_{zz} = \frac{1}{\cos(\theta)} \text{Re}\left[\bm{\lambda}_{\varphi_0} \cdot (\bm{\lambda}_{\text{init}} + \bm{\Gamma} (\bm{I}-\bm{\Gamma})^{-1}\bm{\lambda}_\varphi) \right].
	\end{equation}
The dot product with $\bm{\lambda}_{\varphi_0}$ sums up all of the quasiparticle's possible initial positions.  The vector $\bm{\lambda}_{\text{init}}$ accounts for contributions to conductivity from the time the quasiparticle is created until it reaches an MB junction, while $\bm{\lambda_{\varphi}}$ gives the contribution when the quasiparticle is between MB junctions.  The matrix $\bm{\Gamma}$ describes the connections between orbit segments through both Bragg diffraction and magnetic breakdown, as well as accounting for the exponential damping of the integrand upon traversing a segment of Fermi surface. 

If there are $N$ distinct MB junctions, then $\bm{\lambda}$ are all length-$N$ vectors, $\bm{\Gamma}$ is an $N \times N$ matrix, and $\bm{I}$ is the $N \times N$ identity matrix.  The elements of the $\bm{\lambda}$ vectors are defined as follows:
	\begin{widetext}
	\begin{equation}
	\begin{split}
	\bm{\lambda_{\varphi_0}}[j] &\equiv  \int_{\bm{M}_j}^{\bm{M}_{j+1}} d\varphi_0 \frac{f(\varphi_0)e^{-iG(\varphi_0)}}{\cos\bm{(}\gamma(\varphi_0)\bm{)}} \exp\left(\frac{+1}{\omega_0\tau\cos(\theta)} \int_{\bm{M}_j}^{\varphi_0} \frac{d\varphi'}{\cos\bm{(}\gamma(\varphi')\bm{)}}\right), \\
	\bm{\lambda_{\varphi}}[j] &\equiv \int_{\bm{M}_j}^{\bm{M}_{j+1}} d\varphi \frac{f(\varphi) e^{iG(\varphi)}}{\cos\bm{(}\gamma(\varphi)\bm{)}} \exp \left(\frac{-1}{\omega_0 \tau\cos(\theta)} \int_{\bm{M}_j}^{\varphi} \frac{d\varphi'}{\cos\bm{(}\gamma(\varphi')\bm{)}} \right), \\
	\bm{\lambda_{\text{init}}}[j] &\equiv \int_{\varphi_0}^{\bm{M}_{j+1}} d\varphi \frac{f(\varphi) e^{iG(\varphi)}}{\cos\bm{(}\gamma(\varphi)\bm{)}} \exp \left(\frac{-1}{\omega_0 \tau\cos(\theta)} \int_{\bm{M}_j}^{\varphi} \frac{d\varphi'}{\cos\bm{(}\gamma(\varphi')\bm{)}} \right), \\
	\end{split}
	\end{equation}
	\end{widetext}
where $\bm{M}$ is a vector listing the azimuthal position of all the MB junctions plus a final element that is the first element plus $2\pi$.  The exponentials that appear in these vectors serve to cancel the initial exponential damping of the integrands.

Each row (column) of the matrix $\bm{\Gamma}$ corresponds to a segment of the Fermi surface between MB junctions.  The first row (column) corresponds to the segment between the first two MB junctions, and so on.  Mathematically, this means that the index $i$ corresponds to the segment between $\bm{M}_i$ and $\bm{M}_{i+1}$.  Using this correspondence, we can determine all of the elements of $\bm{\Gamma}$ as follows:
	\begin{equation}\label{eq:Gammadef}
	\bm{\Gamma}_{ij} = 
	\begin{cases}
      	0, \text{\ \ \ if section $i$ has no connection to section $j$;} \\
	\ \\
      	D_i p, \parbox[t]{.37\textwidth}{\ if section $i$ is connected to section $j$ through magnetic breakdown;}\\
	\ \\
	D_i (1-p)\exp\left(-ic\Delta k_{z}^{(i+1 \rightarrow j)}\right), \\ \ \ \ \ \ \ \parbox[t]{.37\textwidth}{if section $i$ is connected to section $j$ through Bragg diffraction.}
    	\end{cases}
	\end{equation}
In the above, $p$ is the probability of magnetic breakdown at a single junction and $D_i$ is the exponential damping of the integrand across segment $i$, given by
	\begin{equation}
	D_i = \exp\left(\frac{-1}{\omega_0\tau\cos(\theta)}\int_{\bm{M}_i}^{\bm{M}_{i+1}} \frac{d\varphi'}{\cos\bm{(}\gamma(\varphi')\bm{)}}\right).
	\end{equation}
The quantity $\Delta k_z^{(i+1 \rightarrow j)}$ is the amount by which $k_z^0$ changes when undergoing Bragg diffraction from the $(i+1)^{\text{th}}$ MB junction to the $j^{\text{th}}$ MB junction.

\section{Results}
\label{sec:results}

We now apply the equations developed in the previous section to underdoped Hg1201.  We consider a simple cosine warping, corresponding to $f(\varphi) = 1$, as well as a staggered twofold warping, corresponding to $f(\varphi) = \sin(2\varphi)$. We consider all of the in-plane Fermi surface shapes described in Section \ref{sec:model}, with their cross-sectional areas fixed to match quantum oscillation measurements~\cite{Barisic2013,Chan2016}.  Each in-plane Fermi surface shape gives us $\gamma(\varphi)$ directly (see Appendix \ref{app:kF}). In addition to the in-plane shape, $G(\varphi)$ is determined by the direction of the magnetic field ($\theta$ and $\phi$) as well as the value of $c$, the interlayer lattice constant.  We use $c = 9.517$ \AA, the value of the interlayer lattice parameter in as-grown Hg1201 with $T_c \approx$ 80 K~\cite{Zhao2006}.  \footnote{The constraint used for the cross-sectional Fermi surface area is from a sample that also had an as-grown $T_c$ of roughly 80 K but that was subsequently heat-treated in a nitrogen-rich atmosphere to achieve a $T_c$ of 74 K~\cite{Chan2016}.  This treatment may slightly decrease the interlayer lattice parameter since it removes interstitial oxygen, but such an effect should be neglible: only a fraction of a percent change, based on similar behavior in YBCO~\cite{Liang2006}.}

We take $\rho_{zz} =\frac{1}{\sigma_{zz}}$, which is justified for a material such as Hg1201 that has $v_z \ll v_x, v_y$~\cite{Prentice2016}.  Since we are only calculating $\sigma_{zz}$ up to a constant of proportionality, we plot all of our results as $\rho_{zz}/\rho_{zz}(\theta = 0)$.

	\begin{figure*}
    	\includegraphics[trim=0.5cm 0cm 0.5cm 1cm, clip, width=\textwidth]{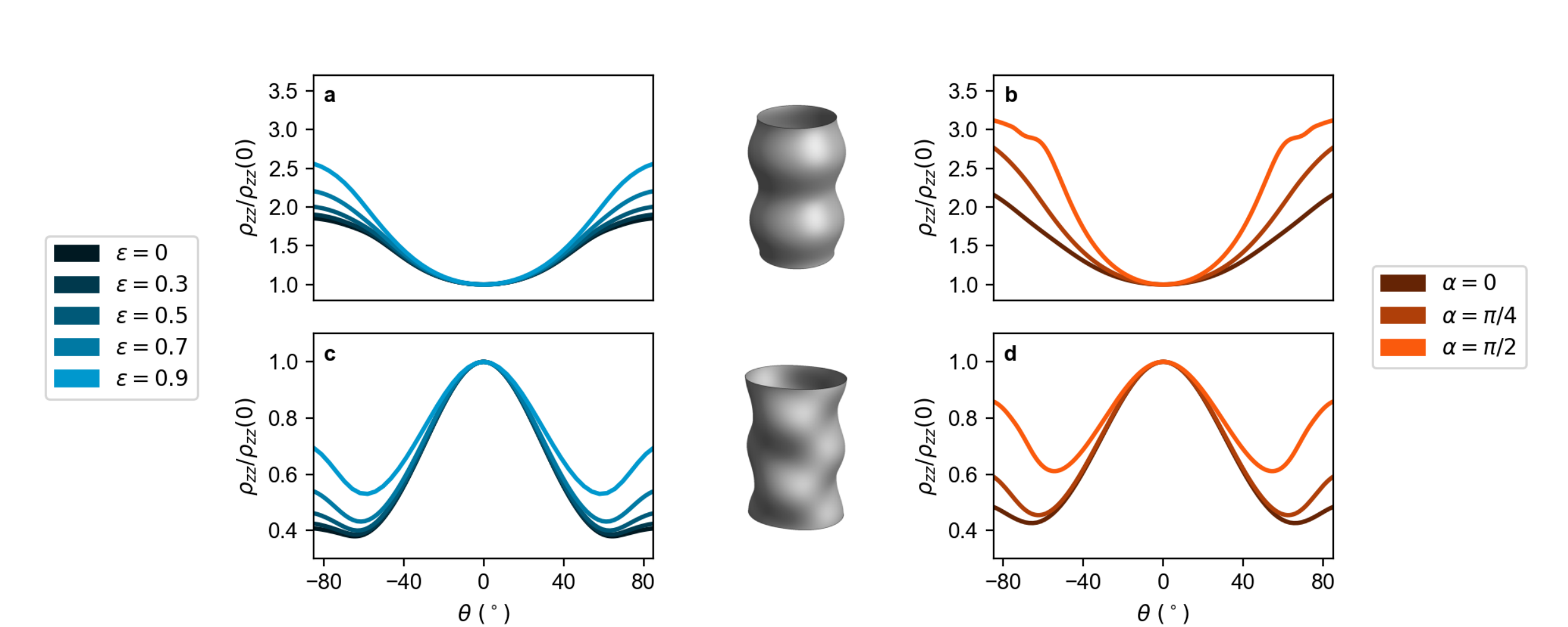}
  	\caption{(Color online) Calculated ADMR of Hg1201 with $\omega_0\tau = 1$ and $\phi = 0$ for various combinations of in-plane shape and interlayer warping.  The top panels show ADMR for a simple cosine interlayer warping with a) elliptical and b) diamond-shaped in-plane Fermi surfaces.  The bottom panels show ADMR for a staggered twofold interlayer warping with c) elliptical and d) diamond-shaped in-plane Fermi surfaces.  While the different in-plane Fermi surfaces cause some variation, there is a clear qualitative distinction between the two interlayer warpings regardless of the in-plane shape.}
  	\label{fig:results1}
	\end{figure*}

\begin{figure*}[]
    	\includegraphics[trim=0.5cm 0cm 0.5cm 0cm, clip, width=\textwidth]{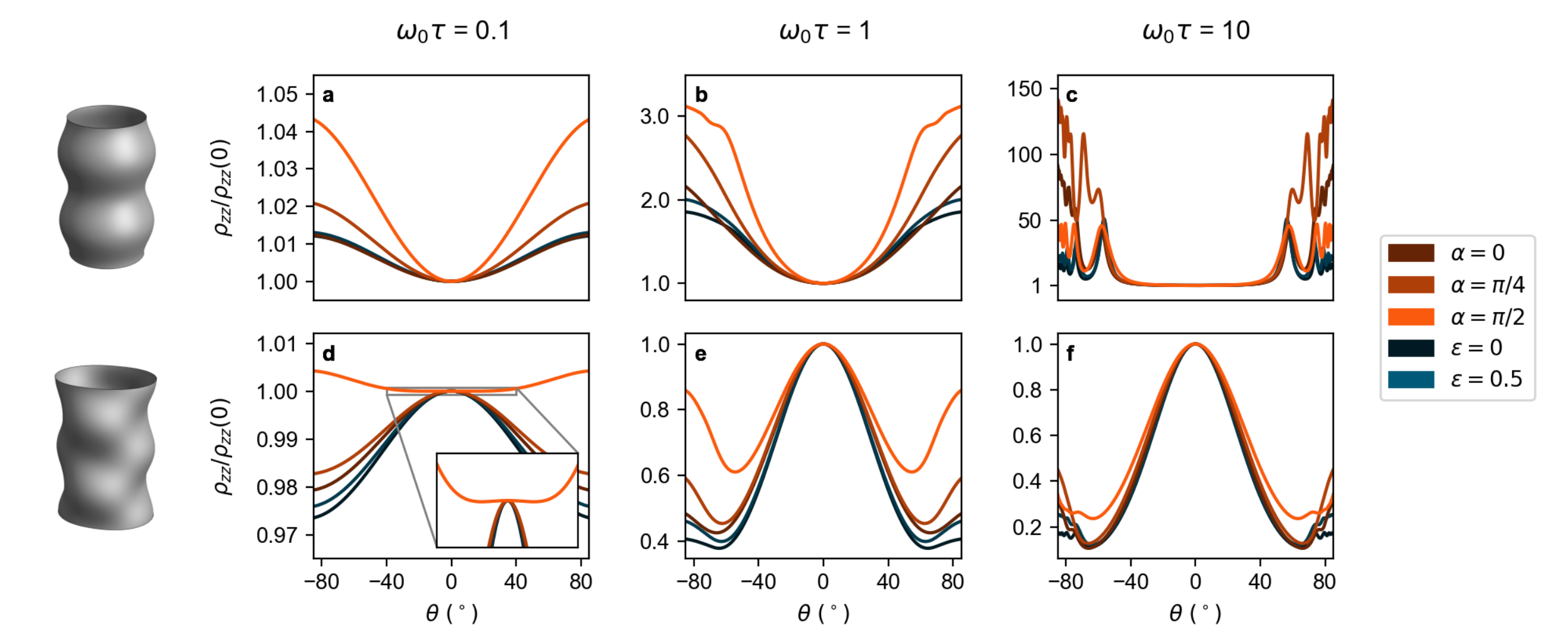}
  	\caption{(Color online) Calculated ADMR of Hg1201 for select in-plane Fermi surface shapes, with $\phi = 0$ and with (a,d) $\omega_0\tau = 0.1$, (b,e) $\omega_0\tau = 1$,  and (c,f) $\omega_0\tau = 10$.  Plots in the top row show ADMR calculated with a simple cosine interlayer warping, while those in the bottom row show ADMR calculated with a staggered twofold interlayer warping.}
  	\label{fig:wtresults}
	\end{figure*}

The main result of this paper is summarized in Fig. \ref{fig:results1}, which shows the expected ADMR for all of the in-plane Fermi surfaces considered (Fig. \ref{fig:FSshapes}) with $\phi=0$ and $\omega_0\tau=1$ and with no Bragg diffraction or magnetic breakdown.  There are quantitative distinctions between the ADMR for different diamond-like and ellipsoidal in-plane Fermi surfaces, but the most notable result is that there is a clear qualitative difference between the two interlayer warpings, even given highly different in-plane Fermi surface geometries:  the simple cosine warping causes a dip in resistance at $\theta = 0$ (i.e., with the magnetic field in the interlayer direction), while the staggered twofold warping yields a hump. This qualitative distinction is not affected by varying $\phi$, as shown in Appendix \ref{app:phidep}, nor does it depend on the value of $\omega_0\tau$.  As shown in Fig. \ref{fig:wtresults}, varying $\omega_0\tau$ from 0.1 to 10 yields the same qualitative difference, and though the distinction becomes less pronounced as $\omega_0\tau$ decreases and $\alpha \rightarrow \pi/2$, it is always present. This result demonstrates that ADMR can be used to unambiguously reveal the interlayer warping for a wide range of scattering times and magnetic fields, in a way that is independent of the in-plane shape.  At higher $\omega_0\tau$ (i.e., at higher magnetic fields and lower temperatures) angle-dependendent magnetoresistance oscillations are apparent, which can be used to discern the in-plane Fermi surface.

A full study of the ADMR of Hg1201 as a function of $\phi$ can be used to gain additional information about the Fermi surface geometry.  As demonstrated in Appendix \ref{app:phidep}, the symmetry of a Fermi surface will be reflected in the symmetry of the ADMR.  The $\phi$-dependence of ADMR is a vital source of information; we do not focus on it in this work simply because its usefuless and interpretation are already well understood~\cite{Hussey2003,Bergemann2003,Ramshaw2017}.

For all of the above, we employ Eq. \ref{eq:Shockley5} to calculate the ADMR of the circular and elliptical in-plane Fermi surfaces, as well as the diamond-shaped pocket with $\alpha = 0$.  The calculation for concave diamond-like pockets ($\alpha > 0$) takes an excessively long time to execute due to the divergent nature of the function $1/\cos\bm{(}\gamma(\varphi)\bm{)}$ at the tips of the diamonds.  For these in-plane Fermi surface shapes, we use the vectorized form of the Shockley tube integral (Eq. \ref{eq:vectorShockley}), which was originally derived for materials exhibiting magnetic breakdown. For the cases considered above, where there is no magnetic breakdown, we choose $B_0/B = 10000$ so that the probability of breakdown approaches 0 and therefore the quasiparticles always trace out the concave diamond-shaped pocket, as shown in Fig. \ref{fig:BraggvsMB}(a).  See Appendix \ref{app:CDWMB} for the mathematical details of the calculations.

\begin{figure*}
    	\includegraphics[trim=0.5cm 0cm 0.5cm 0cm, clip, width=\textwidth]{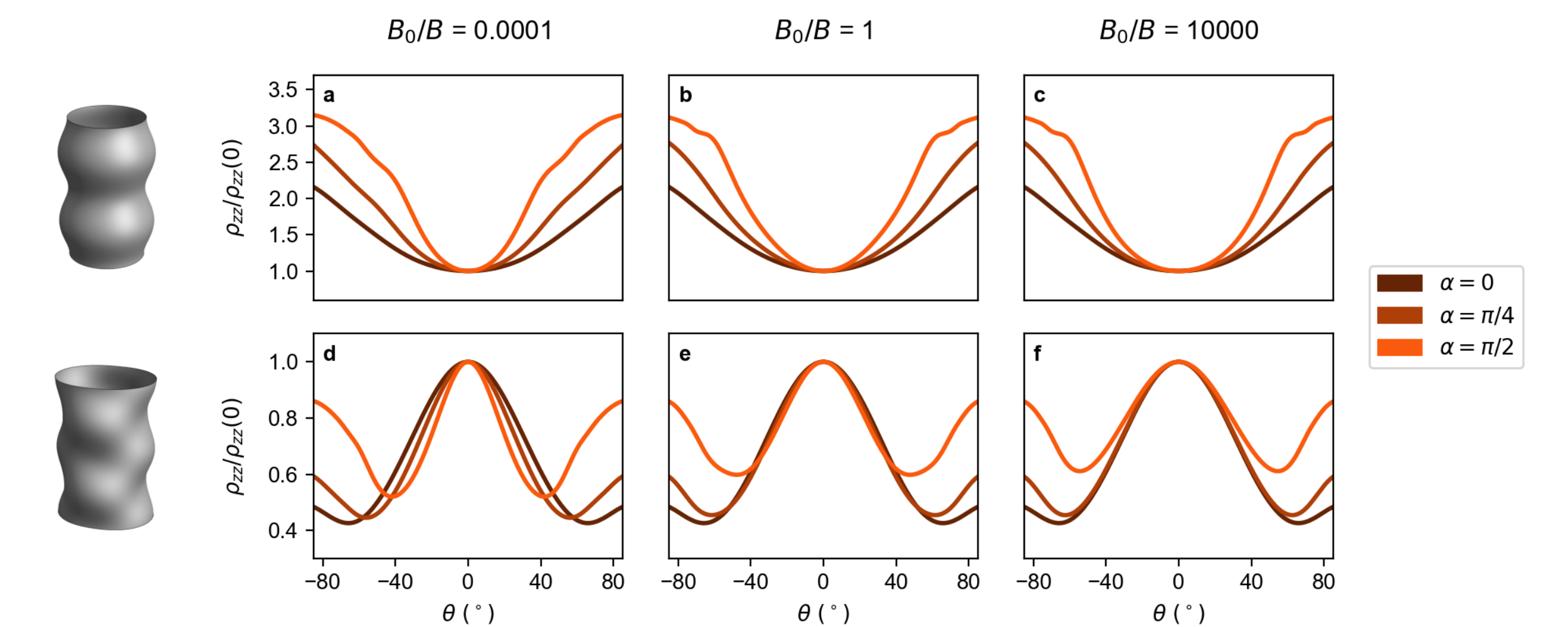}
  	\caption{(Color online) Calculated ADMR of Hg1201 for diamond-shaped in-plane Fermi surfaces with $\phi = 0$ and $\omega_0\tau = 1$ and with various degrees of magnetic breakdown: (a,d) $B_0/B = 0.0001$, (b,e) $B_0/B = 1$,  and (c,f) $B_0/B = 10000$.  Plots in the top row show ADMR calculated with a simple cosine interlayer warping, while those in the bottom row show ADMR calculated with a staggered twofold interlayer warping.}
  	\label{fig:resultswt}
	\end{figure*}

As noted previously, magnetic breakdown effects should be studied when considering the possibility of diamond-shaped pockets being formed by CDW reconstruction.  Therefore, for the diamond-shaped pockets we consider Bragg diffraction and magnetic breakdown effects, using the full force of Eq. \ref{eq:vectorShockley}; see Appendix \ref{app:CDWMB} for details.  For these calculations we again use $\omega_0\tau = 1$ and $\phi = 0$.  The results for three different values of $B_0/B$ are shown in Fig. \ref{fig:resultswt}; with $B_0/B = 0.0001$ there will be near total magnetic breakdown, while $B_0/B = 10000$ leads to effectively zero magnetic breakdown.   It should be noted that quantum oscillation measurements of Hg1201 have set a lower bound of $B_0 \gtrsim 200$ T~\cite{Chan2016}.  Given this, and the fact that $B_0/B = 1$ and $B_0/B = 10000$ produce remarkably similar results, we conclude that magnetic breakdown effects can safely be neglected for this scenario.

We also investigate how the ADMR of Hg1201 would be affected by different periodicities of the interlayer warping.  We repeat our calculations for a circular in-plane Fermi surface but with the interlayer lattice parameter, $c$, reduced by half.  This is equivalent to considering a warping with periodicity $\frac{4\pi}{c}$ rather than $\frac{2\pi}{c}$. The results of these calculations are shown in Appendix \ref{app:periodicity}.  While a change in periodicity will indeed affect the ADMR, it will only be possible to easily distinguish different periodicities at fairly high magnetic fields, such that $\omega_0\tau \gg 1$.

\section{Discussion}
\label{sec:disc}
There is an intuitive way to understand the qualitative distinctions seen in the ADMR between the two different interlayer warpings. As seen from the Shockley-Chambers tube integral, the interlayer conductivity is determined by the correlation of the velocity $v_z$ at angle $\varphi$ with the velocity $v_z$ at $\varphi_0$ on a given Fermi surface orbit, with contributions from all the orbits then summed together.  As shown in Fig. \ref{fig:interlayerwarpings}(a), for $\theta = 0$ the quasiparticles on a Fermi surface that has a simple cosine warping will have a constant $z$-axis component of the Fermi velocity.  This will maximize the correlation of $v_z(\varphi)$ and $v_z(\varphi_0)$ for each quasiparticle orbit, and we expect to see a maximum in conductivity (a minimum in ADMR) at $\theta = 0$.  In contrast, for the staggered twofold warping, $\theta = 0$ gives a minimal correlation of the $z$-axis velocity, so we expect a peak in the ADMR. For Fermi surfaces that have a constant $k_F(\varphi)$, like those in Fig. \ref{fig:interlayerwarpings}, Eq. \ref{eq:Shockley5} can be solved analytically and we can prove that a minimum and maximum, respectively, must occur at $\theta = 0$ for the two interlayer warpings.  Through our numerical calculations, we have shown that the effect is robust to the various in-plane Fermi surface geometries we consider and to magnetic breakdown effects.

\begin{figure}
    	\includegraphics[trim=0.5cm 0.6cm 1cm 0.5cm, clip, width=0.5\textwidth]{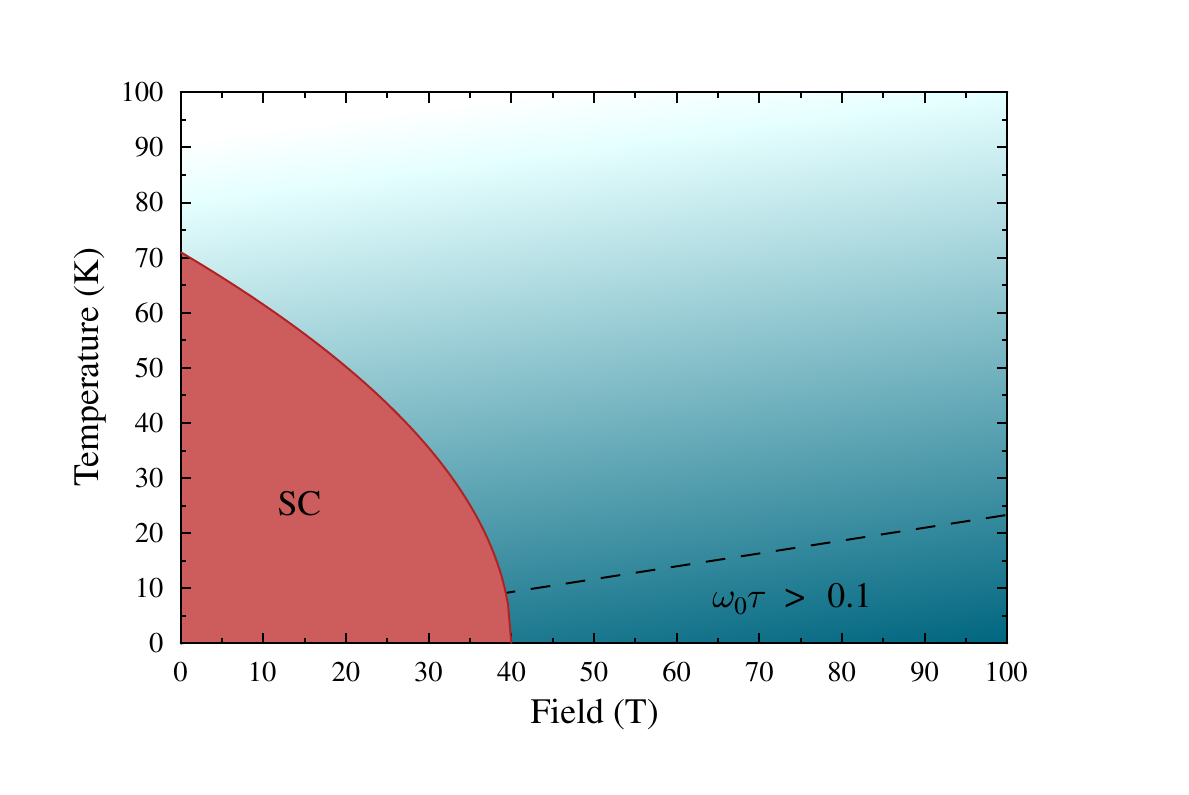}
  	\caption{(Color online) A schematic diagram of where the ADMR measurements described in this paper would be feasible, based on the data and assumptions described in the text.  The dome on the left labeled ``SC" is the superconducting region, based on an estimate of $H_{c2}$ with the magnetic field in the interlayer direction.  Outside the superconducting region, the color gradient represents the value of $\omega_0\tau$, with darker color denoting a larger value.  The dashed line indicates an estimated cutoff for the criterion $\omega_0\tau \geq 0.1$; the region below the dashed line is where the distinction between interlayer warpings could most clearly be distinguished by ADMR measurements.} 
  	\label{fig:measurement}
	\end{figure}

It is worth noting that while this difference decreases as $\omega_0\tau \rightarrow 0$, it should always be evident in a sufficiently accurate experiment.  Assuming $\omega_0\tau = 0.1$ is enough to resolve the effect, we can estimate the temperature and field ranges at which the ADMR should be measured.  Fig. \ref{fig:measurement} shows a schematic diagram of where ADMR measurements are feasible for a sample of Hg1201 with $T_c = 71 K$.  At 45 T and a few Kelvin, Hg1201 has $\omega_c\tau \approx$ 0.35 with the magnetic field perpendicular to the layers ($\theta = 0$), where $\omega_c$ is the cyclotron frequency~\cite{Chan2016}. \footnote{This is the average result across samples with $T_c = 71 K$ and $T_c = 74 K$.}  This gives a bound of $\omega_0\tau \geq 0.35$ for those conditions, since $\omega_c$ is the average of angular velocity $\omega$ about an orbit and therefore $\omega_c \leq \omega_0$ (see Eq. \ref{eq:omegadef}).  We estimate that $\omega_0\tau = 0.35$ at 45 T and 3 K (since Ref. \onlinecite{Chan2016} does not specify the temperature but includes measurements between 1.8 K and 4 K).  The dashed line demarcating $\omega_0\tau = 0.1$ and the gradient representing $\omega_0\tau$ in Fig. \ref{fig:measurement} are based on this estimate and on the approximation that $\tau$ decreases linearly with temperature.  The superconducting dome is drawn by estimating the upper critical field to be $H_{c2} = 40$ T at low temperatures~\cite{Chan2016} and using a shape inspired by the upper critical field evolution of cuprate superconductors in Ref. \onlinecite{Grissonnanche2014}.  For a quasi-two-dimensional material, $H_{c2}(\theta)$ is highly anisotropic~\cite{Bulaevskii1990,Schneider1993,Tinkham1996}, but it will only increase by at most a few percent in the range $-15^\circ < \theta < 15^\circ$.  This angular range should be adequate to resolve staggered or simple cosine warping in the ADMR. 

One caveat in this discussion is that we have assumed the Fermi surface of Hg1201 is a single quasi-two-dimensional pocket.  Quantum oscillation measurements only detect one Fermi surface pocket in Hg1201, but this does not preclude the existence of additional open Fermi surface sheets.  If such sheets are present, they would also contribute to interlayer magnetoresistance, making the interpretation of ADMR measurements more complicated.  While quasi-one-dimensional Fermi surface sheets are difficult to detect directly, open orbits should lead to Lebed minima in the ADMR~\cite{Lebed2004}; if these minima are not observed at high $\omega_0\tau$, this would be evidence that such sheets do not exist or contribute to the magnetoresistance.  The density of states associated with such sheets can also be estimated through careful analysis of the wave form of quantum oscillations~\cite{Harrison1996}.  Such an analysis could be used to determine how important a role, if any, quasi-one-dimensional sheets play in this material.

\section{Conclusion}
\label{sec:conc}

We have calculated the expected interlayer ADMR for underdoped Hg1201 using several possible Fermi surface geometries.  We have shown that the simple cosine and staggered twofold interlayer warpings will have two clearly distinct experimental signatures regardless of whether the in-plane Fermi surface is circular, elliptical, or diamond-shaped.  We have also shown that magnetic breakdown between Fermi surface pockets should not affect these results.  Since ADMR, unlike quantum oscillations, is measured at a fixed field strength, it could be used to study the Fermi surface geometry of Hg1201 at various temperatures and magnetic fields, as illustrated in Fig. \ref{fig:measurement}. Such information on the field- and temperature-dependence of the Fermi surface would be complementary to other experiments that can probe how the ordered states in Hg1201 evolve.  If changes in the Fermi surface geometry can be connected to the evolution of CDW order, or some other ordered state, it would help conclusively answer the question of what drives Fermi surface reconstruction in the underdoped cuprates. 

\section{Acknowledgments}

We thank Mun K. Chan for useful discussions.  S.K.L. acknowledges support from the National Science Foundation Graduate Research Fellowship under Grant No. DGE 1106400.  This research used resources of the National Energy Research Scientific Computing Center, a DOE Office of Science User Facility supported by the Office of Science of the U.S. Department of Energy under Contract No. DE-AC02-05CH11231.  This work was supported by the Gordon and Betty Moore Foundation's EPiQS Initiative through Grant GBMF4374.

\appendix

\section{ADMR for different values of $\phi$}
\label{app:phidep}

	\begin{figure*}[]
    	\includegraphics[trim=0.5cm 0cm 0cm 0cm, clip, width=\textwidth]{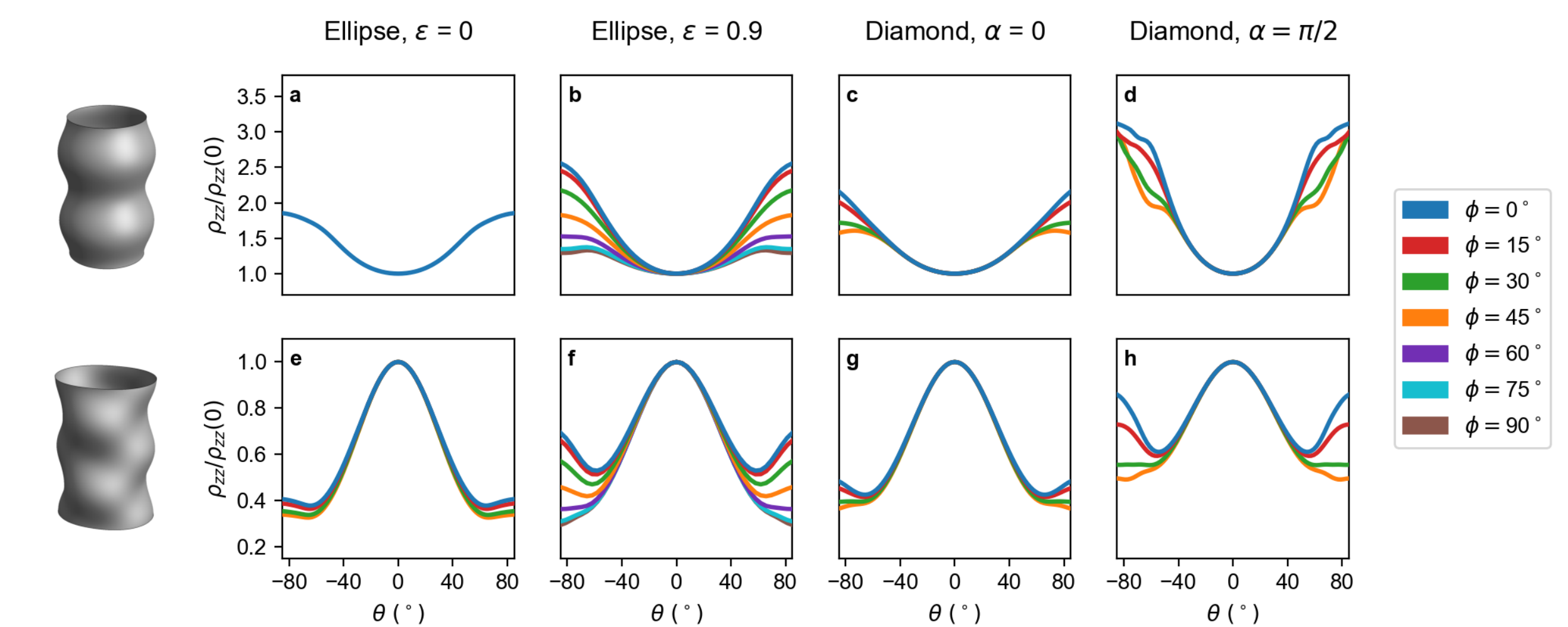}
  	\caption{(Color online) Calculated ADMR of Hg1201 with a simple cosine interlayer warping (top row) or a staggered twofold interlayer warping (bottom row) for 8 different values of $\phi$ as indicated in the legend.  Results are shown for four different in-plane Fermi surfaces: (a,e) circular, (b,f) elliptical, (c,g) square diamond, and (d,h) concave diamond.} 
  	\label{fig:phiresults}
	\end{figure*}

ADMR as a function of $\phi$, the azimuthal angle of the applied field, can provide valuable insights regarding a material's Fermi surface geometry.  We show some examples of ADMR with varying $\phi$ in Fig. \ref{fig:phiresults} to emphasize two points: first, that the symmetry of the Fermi surface determines the symmetry of the ADMR; second, that the qualitative distinction between the two types of interlayer warping considered here is present regardless of the value of $\phi$.

To produce Fig. \ref{fig:phiresults} we consider four different in-plane Fermi surfaces and two different interlayer warpings.  For each combination of these, we calculate ADMR with values of $\phi$ ranging from 0\textdegree\ to 360\textdegree\ in 15\textdegree\ increments.  The results are only shown up to 90\textdegree\ because beyond that, all of the results would lie over curves that are already plotted.  Indeed, for most of the Fermi surfaces considered there is already considerable degeneracy below 90\textdegree: for example,  for the diamond-shaped pockets there is a mirror symmetry about 45\textdegree, and for the circular Fermi surface with simple cosine warping there is complete rotational symmetry.  The symmetry of each Fermi surface is reflected in the $\phi$-dependence of its ADMR, as it must be.

Note that regardless of $\phi$, a Fermi surface with simple cosine warping yields a resistivity minimum at $\theta = 0$, while a Fermi surface with staggered twofold warping yields the opposite.

\section{ADMR for different periodicities of the interlayer warping}
\label{app:periodicity}

Sebastian \textit{et al.} showed that their quantum oscillation data for YBCO could be explained by a  CDW ordering with $l = \nicefrac{1}{2}$ that yields an interlayer warping of periodicity $\frac{2\pi}{c}$.  However, x-ray scattering measurements of the same compound found a three-dimensional CDW order with $l \sim 1$~\cite{Gerber2015}.  If we assume the same type of reconstruction that Sebastian \textit{et al.} suggested, this should lead to an interlayer warping of period $\frac{4\pi}{c}$.  Doubling the periodicity is mathematically equivalent to reducing $c$, the interlayer lattice parameter, by half.

\begin{figure*}
    	\includegraphics[trim=-0.8cm 0cm 0.8cm 0cm, clip, width=\textwidth]{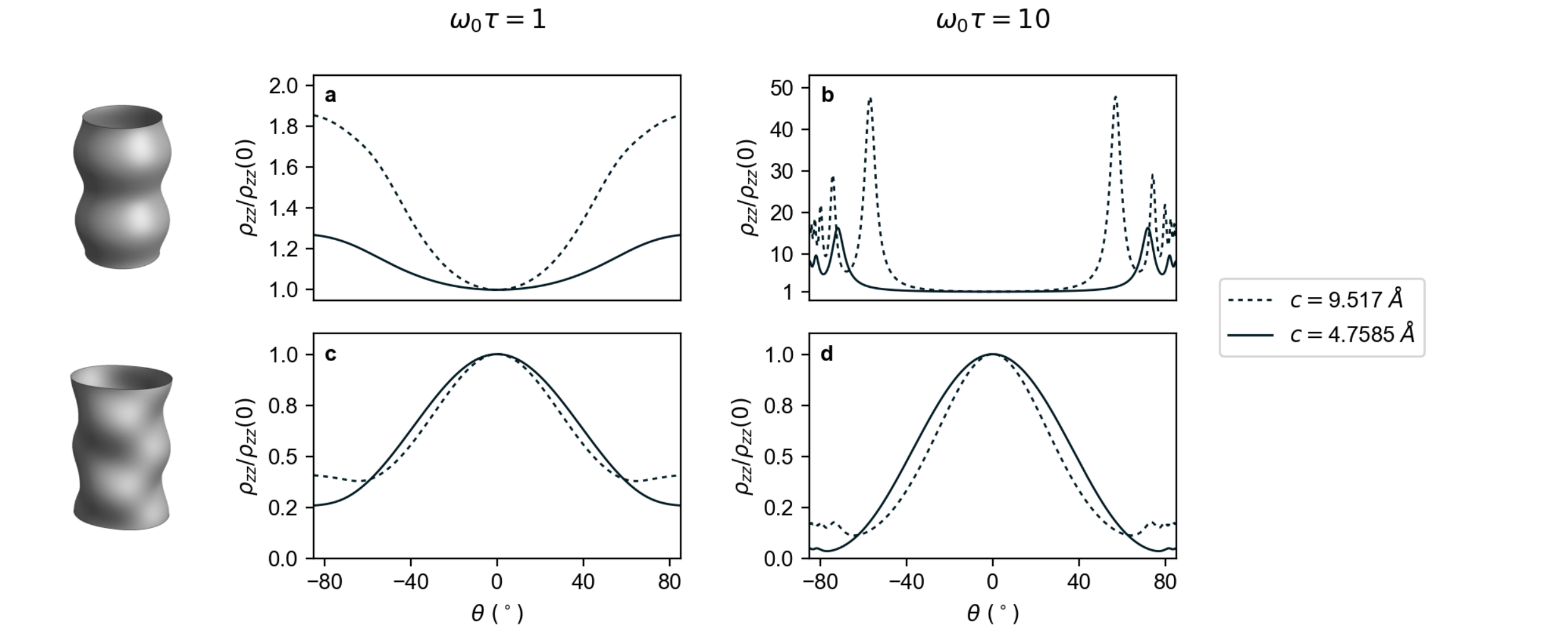}
  	\caption{(Color online) Calculated ADMR of Hg1201 with two different interlayer lattice parameters.   Results are shown for a circular in-plane Fermi surface with a simple cosine interlayer warping (top row) or a staggered twofold interlayer warping (bottom row).  (a,c) $\omega_0\tau = 1$, (b,d) $\omega_0\tau = 10$.} 
  	\label{fig:periodicityresults}
	\end{figure*}

We repeat our calculations for a circular in-plane Fermi surface, but now using $c = 4.7585$ \AA\ rather than the actual interlayer lattice parameter of $c = 9.517$ \AA.  The results are shown in Fig. \ref{fig:periodicityresults}.  Clearly, changing the interlayer lattice parameter produces a quantitative change in the ADMR of Hg1201.  However, given the many unknown parameters with regards to this Fermi surface, such a change may be difficult to distinguish at $\omega_0\tau = 1$; for example, with a simple cosine interlayer warping, halving $c$ produces roughly the same change as increasing the eccentricity of the in-plane Fermi surface (cf. Fig. \ref{fig:results1}).  On the other hand, with $\omega_0\tau = 10$ the ADMR results have more structure, as angle-dependent magnetoresistance oscillations are evident.  Thus it may be possible to use ADMR to determine the periodicity of the interlayer warping, but only if measurements can be taken at very low temperatures and high magnetic fields.

\section{Derivation of $\omega$}
\label{app:omegac}

We define $\omega = \frac{d\varphi}{dt}$, where $\varphi$ is the azimuthal position of a quasiparticle in the $k_x$-$k_y$ plane.  A quasiparticle orbiting around the Fermi surface will have periodic motion in both $k_z$ and $\varphi$.  In the case we are considering, of very small interlayer warping, a change in $k_z$ will have a negligible affect on the Fermi velocity and on the in-plane Fermi momentum.  Therefore, the motion in $k_z$ and in $\varphi$  are separable, and we can base our calculation of $\omega$ on the motion of the quasiparticle about the intralayer Fermi surface.  This motion will be driven by the component of the magnetic field in the interlayer direction, $B\cos(\theta)$.  Since $v_z \ll v_x, v_y$ for this material we can approximate that $v_F$ is entirely perpendicular to the interlayer field.  If we define $k_{\parallel}$ to be the component of momentum along the circumference of the in-plane orbit, then the equation of motion can be written as follows:
	\begin{equation}
	\frac{dk_{\parallel}}{dt} = \frac{e B\cos(\theta)}{\hbar} v_F(\varphi).
	\end{equation}
We find $v_F(\varphi) = \frac{\hbar k_F(\varphi)}{m^*}$ from the dispersion given in Eq. \ref{eq:FS}, so 
	\begin{equation}
	\frac{dk_{\parallel}}{dt} = \frac{e  B\cos(\theta)}{m^*} k_F(\varphi).
	\end{equation}

We want $\frac{d\varphi}{dt}$, so we must now find the relationship between $k_{\parallel}$ and $\varphi$.  This turns out to simply be the equation for an arc-length in polar coordinates.  Noting that $dk_{\parallel}$ is the differential displacement along the orbit and the radius of the in-plane orbit is given by $k_F(\varphi)$, we have

	\begin{equation}
	\frac{dk_{\parallel}}{d\varphi} = \sqrt{k_F(\varphi)^2 + \left(\frac{dk_F(\varphi)}{d\varphi}\right)^2}.
	\end{equation}

Putting everything together, this yields
	\begin{equation}
	\omega =  \frac{d\varphi}{dt} = \frac{d\varphi}{dk_{\parallel}}\frac{dk_{\parallel}}{dt} =  \frac{eB\cos(\theta)}{m^*} \frac{k_F(\varphi)}{\sqrt{k_F(\varphi)^2 + \left(\frac{dk_F(\varphi)}{d\varphi}\right)^2}}.
	\end{equation}
We can simplify the above expression by noting that
	\begin{equation}\label{eq:cosgamma}
	\cos\bm{(}\gamma(\varphi)\bm{)} = \frac{k_F(\varphi)}{\sqrt{k_F(\varphi)^2 + \left(\frac{dk_F(\varphi)}{d\varphi}\right)^2}},
	\end{equation}
where $\gamma(\varphi)$ is the angle between $v_F(\varphi)$ and $k_F(\varphi)$~\cite{Hussey2003a}.  Then we arrive at the form we use in the paper:
	\begin{equation}
	\omega = \frac{eB\cos(\theta)}{m^*}\cos\bm{(}\gamma(\varphi)\bm{)}.
	\end{equation}

\section{The form of the Shockley-Chambers tube integral}
\label{app:Shockley}

When $k_B T \ll E_F$, the Shockley-Chambers tube integral generally appears in the form~\cite{Ziman1972, Yagi1992}

	\begin{equation}\label{eq:Shockleyder2a}
	\begin{split}
		\sigma_{\alpha\beta} = \frac{e^2}{4\pi^3 \hbar^2} & \int dk_H \frac{m_c}{\omega_c} \int_0^{2\pi} d\psi' \int_{0}^{\infty} d\psi \\*
		& v_\alpha(\psi',k_H) v_\beta(\psi'-\psi,k_H) e^{-\psi/\omega_c\tau},
	\end{split}
	\end{equation}
where all velocities are taken to be at the Fermi energy.  The quantity $k_H$ is the component of momentum parallel to the magnetic field and $f_0$ is the equilibrium distribution function.  With a small change of variables, this becomes

	\begin{equation}\label{eq:Shockleyder2b}
	\begin{split}
		\sigma_{\alpha\beta} = \frac{e^2}{4\pi^3 \hbar^2} & \int dk_H \frac{m_c}{\omega_c} \int_0^{2\pi} d\psi' \int_{-\infty}^{\psi'} d\psi'' \\*
		& v_\alpha(\psi',k_H) v_\beta(\psi'',k_H) e^{(\psi''-\psi')/\omega_c\tau}.
	\end{split}
	\end{equation}

It can be shown that, as long as $v_\alpha$ and $v_\beta$ are periodic about the Fermi surface (which they must be), this is equivalent to

	\begin{equation}\label{eq:Shockleyder2c}
	\begin{split}
		\sigma_{\alpha\beta} = \frac{e^2}{4\pi^3 \hbar^2} & \int dk_H \frac{m_c}{\omega_c} \int_0^{2\pi} d\psi' \int_{\psi'}^{\infty} d\psi'' \\*
		& v_\beta(\psi',k_H) v_\alpha(\psi'',k_H) e^{(\psi'-\psi'')/\omega_c\tau}.
	\end{split}
	\end{equation}

In the above equation, $\omega_c$ is not the rate of change of the quasiparticle's azimuthal position.  Rather, it is the cyclotron frequency, defined as $\omega_c \equiv 2\pi/T$ where $T$ is the period of an electron orbit.  By this definition, $\omega_c$ depends on the Fermi surface geometry and the angle of the applied field, but it is constant for a given quasiparticle orbit.  The cyclotron mass $m_c$ is defined as $m_c \equiv eB/\omega_c$; it is not the same as the quasiparticle's effective mass.

The phase variable $\psi'$ in equation \ref{eq:Shockleyder2c} is defined by $d\psi' = \omega_c dt$; by construction, this quantity increases at a constant rate.  If the Fermi surface in question is a perfect cylinder (or sphere), then $\psi'$ is simply the azimuthal position of the quasiparticle.  But in general, a quasiparticle's azimuthal position, which we will call $\varphi'$, will not change at a constant rate about an orbit so $\varphi'$ and $\psi'$ are distinct.  The quasiparticle velocity as a function of $\varphi'$ is simple to obtain geometrically; not so with $\psi'$.  Therefore, we want to rewrite our equation in terms of $\varphi'$ rather than $\psi'$.  

The relationship between the two quantities is given by
	\begin{equation}
	\frac{d\psi'}{d\varphi'} = \frac{d\psi'/dt}{d\varphi'/dt} = \frac{\omega_c}{\omega(\varphi')},
	\end{equation}
where $\omega$ is the azimuthal angular velocity and depends on $\varphi'$.  Note that $\psi'$, just like $\varphi'$, is $0$ at the beginning of an orbit and $2\pi$ after completing an orbit; thus, our limits of integration do not have to change.

Since $\psi''$ varies from $\psi'$ to $\infty$, we can define $\psi'' = \psi' + \omega_c t$, where $t$ varies from $0$ to $\infty$.  Now we have 
	\begin{equation}\label{eq:Shockleyder3}
	\begin{split}
		\sigma_{\alpha\beta} = \frac{e^2}{4\pi^3 \hbar^2} & \int dk_H \frac{m_c}{\omega_c} \int_0^{2\pi} d\varphi' \frac{\omega_c}{\omega(\varphi')} \int_{0}^{\infty} dt\ \omega_c \\*
		& v_\beta(\varphi',k_H) v_\alpha(t,k_H) e^{-t/\tau}.
	\end{split}
	\end{equation}
We can define a variable $\varphi''$ such that $\frac{d\varphi''}{dt} = \omega(\varphi'')$ and $\varphi''(t=0) = \varphi'$, yielding
	\begin{equation}\label{eq:Shockleyder4}
	\begin{split}
		\sigma_{\alpha\beta} = &\frac{e^2}{4\pi^3 \hbar^2} \int dk_H \frac{m_c}{\omega_c} \int_0^{2\pi} d\varphi' \frac{\omega_c}{\omega(\varphi')}  \int_{\varphi'}^{\infty}  d\varphi''\ \frac{\omega_c}{\omega(\varphi'')} \\*
		& v_\beta(\varphi',k_H) v_\alpha(\varphi'',k_H) \exp\left( -\int_{\varphi'}^{\varphi''} \frac{d\varphi'''}{\omega(\varphi''')\tau}\right).
	\end{split}
	\end{equation}

As illustrated in Ref. \onlinecite{Lewin2015a}, we can substitute $k_H = k_z^0 \cos(\theta)$.  We can also substitute $m_c = eB/\omega_c$ and simplify to obtain
	\begin{equation}\label{eq:Shockleyder5}
	\begin{split}
		\sigma_{\alpha\beta} = &\frac{e^3B\cos(\theta)}{4\pi^3 \hbar^2} \int dk_z^0 \int_0^{2\pi} d\varphi'  \frac{ v_\beta(\varphi',k_z^0)}{\omega(\varphi')} \\ &\int_{\varphi'}^{\infty} d\varphi'' \frac{v_\alpha(\varphi'',k_z^0) }{\omega(\varphi'')} \exp\left(-\int_{\varphi'}^{\varphi''} \frac{d\varphi'''}{\omega(\varphi''')\tau}\right).
	\end{split}
	\end{equation}

With the exception of the variable names, this is identical to Eq. \ref{eq:Shockley}.

\section{Mathematical descriptions of the in-plane Fermi surfaces}
\label{app:kF}

In this paper, we consider 8 different in-plane Fermi surfaces: 5 ellipses of various eccentricity and 3 different diamond-shaped pockets.  We need a function $k_F(\varphi)$ that describes each of these Fermi surfaces; this will give us $\gamma(\varphi)$ (through use of Eq. \ref{eq:cosgamma}) and $G(\varphi)$.

The ellipses are defined as follows:
\begin{equation}\label{eq:centeredellipse}
k_F(\varphi) = \frac{1}{\sqrt{(\cos(\varphi)/a)^2 + (\sin(\varphi)/b)^2}},
\end{equation}
where $a$ and $b$ are the major and minor axes of the ellipse, respectively.  The area of an ellipse is given by $A = \pi a b$, and the eccentricity is defined as $\epsilon = \sqrt{1-(b/a)^2}$.  In our calculations, we use the cross-sectional Fermi surface area that has been determined by quantum oscillation measurements; therefore, for every $\epsilon$ we consider it is straightforward to calculate the appropriate $a$ and $b$.  Note that Eq. \ref{eq:centeredellipse} describes an ellipse that is centered at the origin, which is necessary for the interlayer warping to be properly imposed on the in-plane Fermi surface.

The diamond-shaped pockets have a more complicated definition, given as follows:

\begin{equation}\label{eq:diamond1}
\begin{split}
k_F(\varphi) =  \frac{R}{2\sin(\beta)} \Bigl((\cos(\beta)+\sin(\beta))(\abs{\cos(\varphi)}+\abs{\sin(\varphi)}) \\*
 - \sqrt{2+(2\cos(\beta)\sin(\beta)+1)(2\abs{\sin(\varphi)\cos(\varphi)}-1)}\Bigr).
\end{split}
\end{equation}

In the above equation, $\beta \equiv \alpha/2$, where $\alpha$ is the angle that must be subtended on a circle to make each arc that forms the diamond-shaped pocket; in this study we consider $\alpha = 0$, $\pi/4$, and $\pi/2$.  The constant $R$ is the distance from the center of the pocket to the tip of each arc.  In order to constrain the Fermi surface area to be a certain $A_{FS}$, we must set $R$ as follows:

\begin{widetext}
\begin{equation}\label{eq:defR}
R = \sqrt{\frac{2A_{FS}\sin^2(\beta)}{\int_0^{\pi/2}\left((\cos(\beta) + \sin(\beta))(\cos(\varphi) + \sin(\varphi)) -  \sqrt{2 +  (2\cos(\beta)\sin(\beta) + 1)(2\sin(\varphi)\cos(\varphi) - 1)}\right)^2 d\varphi}}.
\end{equation}
\end{widetext}

We use a different equation for the diamond-shaped pockets when considering magnetic breakdown effects (that is, when using the vectorized form of the Shockley tube integral).  As shown in Fig. \ref{fig:BraggvsMB}, in this scenario the origin (the center of the Brillouin zone) is in the middle of the 4 pockets, so that the arcs appear convex instead of concave.  In this case, $R$ is still defined by Eq. \ref{eq:defR} (since it is still the area of the concave pockets that is constrained), but we use

\begin{equation}\label{eq:diamond2}
\begin{split}
k_F(\varphi) = \frac{R}{2\sin(\beta)} \Bigl((-\cos(\beta)+ \sin(\beta))(\abs{\cos(\varphi)}+\abs{\sin(\varphi)}) \\*
+ \sqrt{2+(1-2\cos(\beta)\sin(\beta))(2\abs{\sin(\varphi)\cos(\varphi)}-1)} \Bigr).
\end{split}
\end{equation}

Equations \ref{eq:diamond1} and \ref{eq:diamond2} actually diverge for $\alpha = 0$; that is, for a square pocket.  For that case, we use the simpler equation
\begin{equation}
k_F(\varphi) = \frac{\sqrt{A_{FS}/2}}{\abs{\cos(\varphi)}+\abs{\sin(\varphi)}}.
\end{equation}

This equation is valid both for a Fermi surface contained within the Brillouin zone or for pockets in the four corners of the Brillouin zone, since tiling four tilted squares in the corners produces a square of the same size in the center.

The diamond-shaped pockets we consider all have their tips pointing along $k_x$ and $k_y$, which is the orientation we would expect if they arise from a CDW reconstruction~\cite{Chan2016}. The orientation of the staggered twofold warping with respect to the diamonds is based on the results of Ref. \onlinecite{Sebastian2014}.

\begin{figure}[]
    	\includegraphics[trim=0.8cm 0cm 12cm 1.2cm, clip, width=0.5\textwidth]{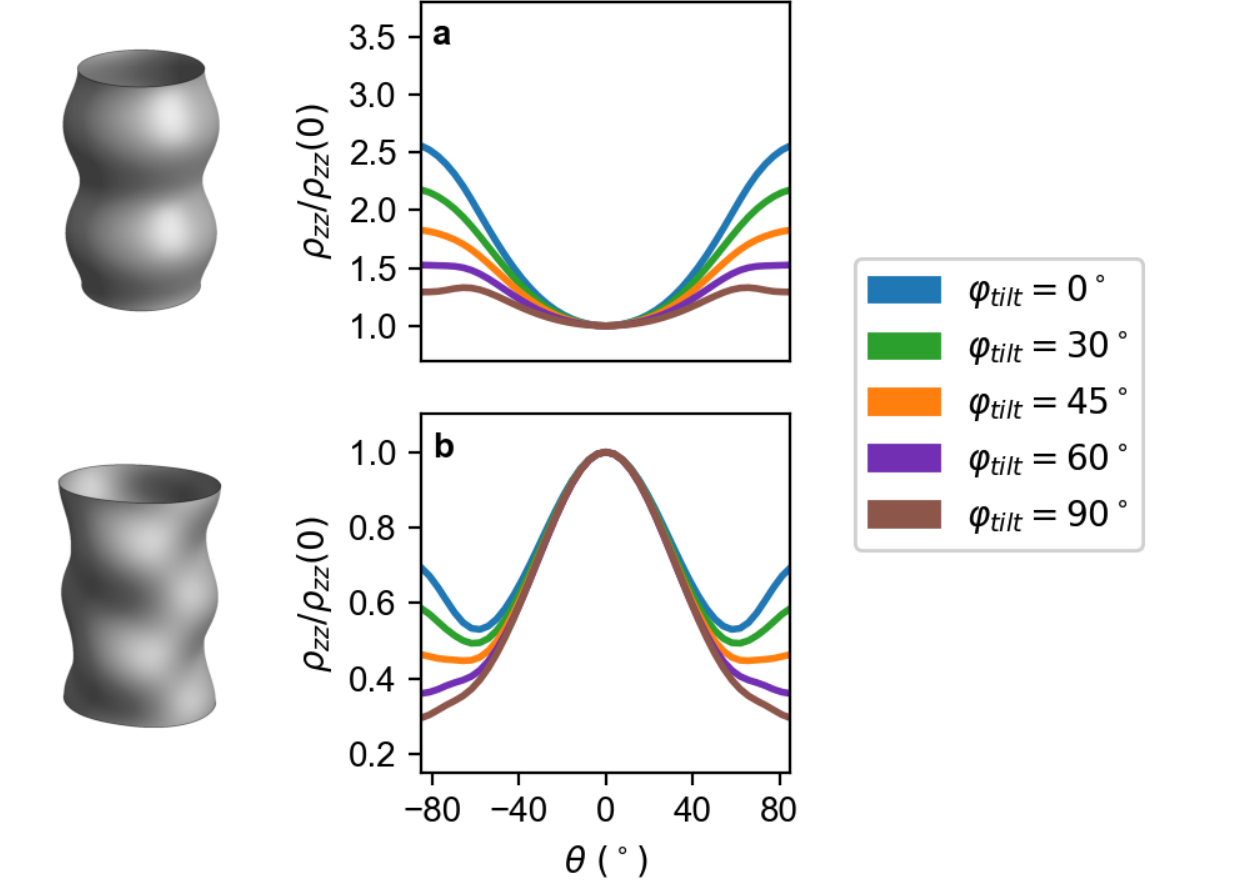}
  	\caption{(Color online) Calculated ADMR of Hg1201 for an elliptical Fermi surface with various orientations.  The in-plane Fermi surface is an ellipse with $\epsilon = 0.9$.  The variable $\varphi_{\text{tilt}}$ indicates the angle between $k_x$ and the major axis of the ellipse.  The magnetic field is fixed at $\phi = 0$ for all calculations.  a) Results for a simple cosine interlayer warping.  b) Results for a staggered twofold interlayer warping.}
  	\label{fig:tiltresults}
	\end{figure}

The ellipses we consider all have their major axes along $k_x$ and minor axes along $k_y$.  But if the Fermi surface pockets are elliptical, they could very well be tilted off-axis; this is illustrated, for instance, in Ref. \onlinecite{Doiron-Leyraud2007}.  For the simple cosine warping, tilting the ellipses in the $k_x-k_y$ plane is simply a matter of rotating the whole Fermi surface, since the interlayer warping has no azimuthal dependence.  But for the staggered twofold warping, the alignment between the in-plane shape and interlayer warping could affect the magnetoresistance in a non-trivial way.  To verify that this will not affect our general conclusions, we calculate ADMR for an ellipse with its major axes tilted away from $k_x$ by some amount $\varphi_{\text{tilt}}$.  We use an ellipse with $\epsilon = 0.9$ and fix the azimuthal angle of the magnetic field to $\phi = 0$.  The results are shown in Fig. \ref{fig:tiltresults}.  With the simple cosine warping, changing the orientation of the in-plane ellipse simply rotates the entire Fermi surface.  Therefore, it is equivalent to rotating the applied magnetic field to a different value of $\phi$; this can be seen by comparing Fig. \ref{fig:tiltresults}(a) to Fig. \ref{fig:phiresults}(b).  For the staggered twofold warping, changing the orientation of the in-plane ellipse does modify the overall magnetoresistance beyond a simple rotation; however, a comparison between Fig. \ref{fig:tiltresults}(b) and Fig. \ref{fig:phiresults}(f) reveals that the change is minimal and does not alter the conclusions of this paper.

\section{Explicit forms of $\bm{M}$, $\bm{\Gamma}$, and $\Delta k_z$ for diamond pocket magnetic breakdown calculations}
\label{app:CDWMB}

The definitions of $\bm{M}$, $\bm{\Gamma}$, and $\Delta k_z$ all depend on the direction of quasiparticle motion.  In this work, we always consider quasiparticles that are moving counterclockwise about the Fermi surface.

In the case of Fermi surface pockets in the four corners of the Brillouin zone, as shown in Fig. \ref{fig:BraggvsMB}, the vector $\bm{M}$ of MB junctions will be

\begin{equation}
\bm{M} = \left[0,\frac{\pi}{2},\pi, \frac{3\pi}{2},2\pi \right].
\end{equation}

For a quasiparticle that is traveling counterclockwise on the Fermi surface, the matrix $\bm{\Gamma}$ will be as follows:

	\begin{equation}\label{eq:Gammaspecific}
	\bm{\Gamma} \equiv D
	\bpm 0 & p & 0 & qe^{-i c \Delta k_{z}^{(2 \rightarrow 4)}} \\
	qe^{-i c \Delta k_z^{(3 \rightarrow 1)}} & 0 & p & 0 \\
	0 & qe^{-i c \Delta k_z^{(4 \rightarrow 2)}} & 0 & p\\
	p & 0 & qe^{-i c\Delta k_z^{(5 \rightarrow 3)}} & 0 \epm,
	\end{equation}
where $p = e^{-B_0/B\cos(\theta)}$ is the magnetic breakdown probability, we have defined $q = 1-p$, and we have defined $D \equiv \exp\left( -\int_0^{\pi/2} \frac{d\varphi'}{\omega(\varphi')\tau}\right)$.  We are able to use a single prefactor of $D$ rather than including $D_i$ in the matrix elements (as in Eq. \ref{eq:Gammadef}) because in this case $\int_{\bm{M}_i}^{\bm{M}_{i+1}} \frac{d\varphi'}{\omega(\varphi')\tau}$ is the same for all $i$.  This is due to the 4-fold symmetry of the in-plane Fermi surface and magnetic breakdown junctions.

In the matrix above, $\Delta k_z^{(i+1 \rightarrow j)}$ is the amount by which $k_z^0$ changes when a quasiparticle undergoes Bragg diffraction from junction $i+1$ to junction $j$.  This can be found geometrically as explained in Ref. \onlinecite{Lewin2015a}:
	\begin{equation}
	\begin{split}
	\Delta k_z^{(i+1 \rightarrow j)} = \tan(\theta)[&k_F(\varphi_j)\cos(\varphi_j - \phi) - \\ &k_F(\varphi_{i+1})\cos(\varphi_{i+1} - \phi)],
	\end{split}
	\end{equation}
where $\varphi_j = \bm{M}[j]$ and $\varphi_{i+1}= \bm{M}[i+1]$.  Note that in writing $\Delta k_z$ in Eq. \ref{eq:Gammaspecific} we have used 1-indexing for $\bm{M}$, e.g. $\bm{M}[1] = 0$, $\bm{M}[2] = \frac{\pi}{2}$, etc.


\begin{thebibliography}{43}%
\makeatletter
\providecommand \@ifxundefined [1]{%
 \@ifx{#1\undefined}
}%
\providecommand \@ifnum [1]{%
 \ifnum #1\expandafter \@firstoftwo
 \else \expandafter \@secondoftwo
 \fi
}%
\providecommand \@ifx [1]{%
 \ifx #1\expandafter \@firstoftwo
 \else \expandafter \@secondoftwo
 \fi
}%
\providecommand \natexlab [1]{#1}%
\providecommand \enquote  [1]{``#1''}%
\providecommand \bibnamefont  [1]{#1}%
\providecommand \bibfnamefont [1]{#1}%
\providecommand \citenamefont [1]{#1}%
\providecommand \href@noop [0]{\@secondoftwo}%
\providecommand \href [0]{\begingroup \@sanitize@url \@href}%
\providecommand \@href[1]{\@@startlink{#1}\@@href}%
\providecommand \@@href[1]{\endgroup#1\@@endlink}%
\providecommand \@sanitize@url [0]{\catcode `\\12\catcode `\$12\catcode
  `\&12\catcode `\#12\catcode `\^12\catcode `\_12\catcode `\%12\relax}%
\providecommand \@@startlink[1]{}%
\providecommand \@@endlink[0]{}%
\providecommand \url  [0]{\begingroup\@sanitize@url \@url }%
\providecommand \@url [1]{\endgroup\@href {#1}{\urlprefix }}%
\providecommand \urlprefix  [0]{URL }%
\providecommand \Eprint [0]{\href }%
\providecommand \doibase [0]{http://dx.doi.org/}%
\providecommand \selectlanguage [0]{\@gobble}%
\providecommand \bibinfo  [0]{\@secondoftwo}%
\providecommand \bibfield  [0]{\@secondoftwo}%
\providecommand \translation [1]{[#1]}%
\providecommand \BibitemOpen [0]{}%
\providecommand \bibitemStop [0]{}%
\providecommand \bibitemNoStop [0]{.\EOS\space}%
\providecommand \EOS [0]{\spacefactor3000\relax}%
\providecommand \BibitemShut  [1]{\csname bibitem#1\endcsname}%
\let\auto@bib@innerbib\@empty
\bibitem [{\citenamefont {Doiron-Leyraud}\ \emph {et~al.}(2007)\citenamefont
  {Doiron-Leyraud}, \citenamefont {Proust}, \citenamefont {LeBoeuf},
  \citenamefont {Levallois}, \citenamefont {Bonnemaison}, \citenamefont
  {Liang}, \citenamefont {Bonn}, \citenamefont {Hardy},\ and\ \citenamefont
  {Taillefer}}]{Doiron-Leyraud2007}%
  \BibitemOpen
  \bibfield  {author} {\bibinfo {author} {\bibfnamefont {N.}~\bibnamefont
  {Doiron-Leyraud}}, \bibinfo {author} {\bibfnamefont {C.}~\bibnamefont
  {Proust}}, \bibinfo {author} {\bibfnamefont {D.}~\bibnamefont {LeBoeuf}},
  \bibinfo {author} {\bibfnamefont {J.}~\bibnamefont {Levallois}}, \bibinfo
  {author} {\bibfnamefont {J.-B.}\ \bibnamefont {Bonnemaison}}, \bibinfo
  {author} {\bibfnamefont {R.}~\bibnamefont {Liang}}, \bibinfo {author}
  {\bibfnamefont {D.~A.}\ \bibnamefont {Bonn}}, \bibinfo {author}
  {\bibfnamefont {W.~N.}\ \bibnamefont {Hardy}}, \ and\ \bibinfo {author}
  {\bibfnamefont {L.}~\bibnamefont {Taillefer}},\ }\href {\doibase
  10.1038/nature05872} {\bibfield  {journal} {\bibinfo  {journal} {Nature}\
  }\textbf {\bibinfo {volume} {447}},\ \bibinfo {pages} {565} (\bibinfo {year}
  {2007})},\ \Eprint {http://arxiv.org/abs/0801.1281} {arXiv:0801.1281}
  \BibitemShut {NoStop}%
\bibitem [{\citenamefont {Bangura}\ \emph {et~al.}(2008)\citenamefont
  {Bangura}, \citenamefont {Fletcher}, \citenamefont {Carrington},
  \citenamefont {Levallois}, \citenamefont {Nardone}, \citenamefont {Vignolle},
  \citenamefont {Heard}, \citenamefont {Doiron-Leyraud}, \citenamefont
  {LeBoeuf}, \citenamefont {Taillefer}, \citenamefont {Adachi}, \citenamefont
  {Proust},\ and\ \citenamefont {Hussey}}]{Bangura_YBCO_2008}%
  \BibitemOpen
  \bibfield  {author} {\bibinfo {author} {\bibfnamefont {A.~F.}\ \bibnamefont
  {Bangura}}, \bibinfo {author} {\bibfnamefont {J.~D.}\ \bibnamefont
  {Fletcher}}, \bibinfo {author} {\bibfnamefont {A.}~\bibnamefont
  {Carrington}}, \bibinfo {author} {\bibfnamefont {J.}~\bibnamefont
  {Levallois}}, \bibinfo {author} {\bibfnamefont {M.}~\bibnamefont {Nardone}},
  \bibinfo {author} {\bibfnamefont {B.}~\bibnamefont {Vignolle}}, \bibinfo
  {author} {\bibfnamefont {P.~J.}\ \bibnamefont {Heard}}, \bibinfo {author}
  {\bibfnamefont {N.}~\bibnamefont {Doiron-Leyraud}}, \bibinfo {author}
  {\bibfnamefont {D.}~\bibnamefont {LeBoeuf}}, \bibinfo {author} {\bibfnamefont
  {L.}~\bibnamefont {Taillefer}}, \bibinfo {author} {\bibfnamefont
  {S.}~\bibnamefont {Adachi}}, \bibinfo {author} {\bibfnamefont
  {C.}~\bibnamefont {Proust}}, \ and\ \bibinfo {author} {\bibfnamefont {N.~E.}\
  \bibnamefont {Hussey}},\ }\href {\doibase 10.1103/PhysRevLett.100.047004}
  {\bibfield  {journal} {\bibinfo  {journal} {Phys. Rev. Lett.}\ }\textbf
  {\bibinfo {volume} {100}},\ \bibinfo {pages} {047004} (\bibinfo {year}
  {2008})}\BibitemShut {NoStop}%
\bibitem [{\citenamefont {Yelland}\ \emph {et~al.}(2008)\citenamefont
  {Yelland}, \citenamefont {Singleton}, \citenamefont {Mielke}, \citenamefont
  {Harrison}, \citenamefont {Balakirev}, \citenamefont {Dabrowski},\ and\
  \citenamefont {Cooper}}]{Yelland_YBCO_2008}%
  \BibitemOpen
  \bibfield  {author} {\bibinfo {author} {\bibfnamefont {E.~A.}\ \bibnamefont
  {Yelland}}, \bibinfo {author} {\bibfnamefont {J.}~\bibnamefont {Singleton}},
  \bibinfo {author} {\bibfnamefont {C.~H.}\ \bibnamefont {Mielke}}, \bibinfo
  {author} {\bibfnamefont {N.}~\bibnamefont {Harrison}}, \bibinfo {author}
  {\bibfnamefont {F.~F.}\ \bibnamefont {Balakirev}}, \bibinfo {author}
  {\bibfnamefont {B.}~\bibnamefont {Dabrowski}}, \ and\ \bibinfo {author}
  {\bibfnamefont {J.~R.}\ \bibnamefont {Cooper}},\ }\href {\doibase
  10.1103/PhysRevLett.100.047003} {\bibfield  {journal} {\bibinfo  {journal}
  {Phys. Rev. Lett.}\ }\textbf {\bibinfo {volume} {100}},\ \bibinfo {pages}
  {047003} (\bibinfo {year} {2008})}\BibitemShut {NoStop}%
\bibitem [{\citenamefont {Bari{\v{s}}i{\'{c}}}\ \emph
  {et~al.}(2013)\citenamefont {Bari{\v{s}}i{\'{c}}}, \citenamefont {Badoux},
  \citenamefont {Chan}, \citenamefont {Dorow}, \citenamefont {Tabis},
  \citenamefont {Vignolle}, \citenamefont {Yu}, \citenamefont {B{\'{e}}ard},
  \citenamefont {Zhao}, \citenamefont {Proust},\ and\ \citenamefont
  {Greven}}]{Barisic2013}%
  \BibitemOpen
  \bibfield  {author} {\bibinfo {author} {\bibfnamefont {N.}~\bibnamefont
  {Bari{\v{s}}i{\'{c}}}}, \bibinfo {author} {\bibfnamefont {S.}~\bibnamefont
  {Badoux}}, \bibinfo {author} {\bibfnamefont {M.~K.}\ \bibnamefont {Chan}},
  \bibinfo {author} {\bibfnamefont {C.}~\bibnamefont {Dorow}}, \bibinfo
  {author} {\bibfnamefont {W.}~\bibnamefont {Tabis}}, \bibinfo {author}
  {\bibfnamefont {B.}~\bibnamefont {Vignolle}}, \bibinfo {author}
  {\bibfnamefont {G.}~\bibnamefont {Yu}}, \bibinfo {author} {\bibfnamefont
  {J.}~\bibnamefont {B{\'{e}}ard}}, \bibinfo {author} {\bibfnamefont
  {X.}~\bibnamefont {Zhao}}, \bibinfo {author} {\bibfnamefont {C.}~\bibnamefont
  {Proust}}, \ and\ \bibinfo {author} {\bibfnamefont {M.}~\bibnamefont
  {Greven}},\ }\href {\doibase 10.1038/nphys2792} {\bibfield  {journal}
  {\bibinfo  {journal} {Nature Physics}\ }\textbf {\bibinfo {volume} {9}},\
  \bibinfo {pages} {761} (\bibinfo {year} {2013})}\BibitemShut {NoStop}%
\bibitem [{\citenamefont {Chan}\ \emph {et~al.}(2016)\citenamefont {Chan},
  \citenamefont {Harrison}, \citenamefont {McDonald}, \citenamefont {Ramshaw},
  \citenamefont {Modic}, \citenamefont {Bari{\v{s}}i{\'{c}}},\ and\
  \citenamefont {Greven}}]{Chan2016}%
  \BibitemOpen
  \bibfield  {author} {\bibinfo {author} {\bibfnamefont {M.~K.}\ \bibnamefont
  {Chan}}, \bibinfo {author} {\bibfnamefont {N.}~\bibnamefont {Harrison}},
  \bibinfo {author} {\bibfnamefont {R.~D.}\ \bibnamefont {McDonald}}, \bibinfo
  {author} {\bibfnamefont {B.~J.}\ \bibnamefont {Ramshaw}}, \bibinfo {author}
  {\bibfnamefont {K.~A.}\ \bibnamefont {Modic}}, \bibinfo {author}
  {\bibfnamefont {N.}~\bibnamefont {Bari{\v{s}}i{\'{c}}}}, \ and\ \bibinfo
  {author} {\bibfnamefont {M.}~\bibnamefont {Greven}},\ }\href {\doibase
  10.1038/ncomms12244} {\bibfield  {journal} {\bibinfo  {journal} {Nature
  Communications}\ }\textbf {\bibinfo {volume} {7}},\ \bibinfo {pages} {12244}
  (\bibinfo {year} {2016})},\ \Eprint {http://arxiv.org/abs/1606.02772}
  {arXiv:1606.02772} \BibitemShut {NoStop}%
\bibitem [{\citenamefont {Ghiringhelli}\ \emph {et~al.}(2012)\citenamefont
  {Ghiringhelli}, \citenamefont {Tacon}, \citenamefont {Minola}, \citenamefont
  {Blanco-Canosa}, \citenamefont {Mazzoli}, \citenamefont {Brookes},
  \citenamefont {{De Luca}}, \citenamefont {Frano}, \citenamefont {Hawthorn},
  \citenamefont {He}, \citenamefont {Loew}, \citenamefont {{Moretti Sala}},
  \citenamefont {Peets}, \citenamefont {Salluzzo}, \citenamefont {Schierle},
  \citenamefont {Sutarto}, \citenamefont {Sawatzky}, \citenamefont {Weschke},
  \citenamefont {Keimer},\ and\ \citenamefont
  {Braicovich}}]{Ghiringhelli2012a}%
  \BibitemOpen
  \bibfield  {author} {\bibinfo {author} {\bibfnamefont {G.}~\bibnamefont
  {Ghiringhelli}}, \bibinfo {author} {\bibfnamefont {M.~L.}\ \bibnamefont
  {Tacon}}, \bibinfo {author} {\bibfnamefont {M.}~\bibnamefont {Minola}},
  \bibinfo {author} {\bibfnamefont {S.}~\bibnamefont {Blanco-Canosa}}, \bibinfo
  {author} {\bibfnamefont {C.}~\bibnamefont {Mazzoli}}, \bibinfo {author}
  {\bibfnamefont {N.~B.}\ \bibnamefont {Brookes}}, \bibinfo {author}
  {\bibfnamefont {G.~M.}\ \bibnamefont {{De Luca}}}, \bibinfo {author}
  {\bibfnamefont {A.}~\bibnamefont {Frano}}, \bibinfo {author} {\bibfnamefont
  {D.~G.}\ \bibnamefont {Hawthorn}}, \bibinfo {author} {\bibfnamefont
  {F.}~\bibnamefont {He}}, \bibinfo {author} {\bibfnamefont {T.}~\bibnamefont
  {Loew}}, \bibinfo {author} {\bibfnamefont {M.}~\bibnamefont {{Moretti
  Sala}}}, \bibinfo {author} {\bibfnamefont {D.~C.}\ \bibnamefont {Peets}},
  \bibinfo {author} {\bibfnamefont {M.}~\bibnamefont {Salluzzo}}, \bibinfo
  {author} {\bibfnamefont {E.}~\bibnamefont {Schierle}}, \bibinfo {author}
  {\bibfnamefont {R.}~\bibnamefont {Sutarto}}, \bibinfo {author} {\bibfnamefont
  {G.~A.}\ \bibnamefont {Sawatzky}}, \bibinfo {author} {\bibfnamefont
  {E.}~\bibnamefont {Weschke}}, \bibinfo {author} {\bibfnamefont
  {B.}~\bibnamefont {Keimer}}, \ and\ \bibinfo {author} {\bibfnamefont
  {L.}~\bibnamefont {Braicovich}},\ }\href@noop {} {\bibfield  {journal}
  {\bibinfo  {journal} {Science}\ }\textbf {\bibinfo {volume} {337}},\ \bibinfo
  {pages} {821} (\bibinfo {year} {2012})}\BibitemShut {NoStop}%
\bibitem [{\citenamefont {Chang}\ \emph {et~al.}(2012)\citenamefont {Chang},
  \citenamefont {Blackburn}, \citenamefont {Holmes}, \citenamefont
  {Christensen}, \citenamefont {Larsen}, \citenamefont {Mesot}, \citenamefont
  {Liang}, \citenamefont {Bonn}, \citenamefont {Hardy}, \citenamefont
  {Watenphul}, \citenamefont {Zimmermann}, \citenamefont {Forgan},\ and\
  \citenamefont {Hayden}}]{Chang2012}%
  \BibitemOpen
  \bibfield  {author} {\bibinfo {author} {\bibfnamefont {J.}~\bibnamefont
  {Chang}}, \bibinfo {author} {\bibfnamefont {E.}~\bibnamefont {Blackburn}},
  \bibinfo {author} {\bibfnamefont {A.~T.}\ \bibnamefont {Holmes}}, \bibinfo
  {author} {\bibfnamefont {N.~B.}\ \bibnamefont {Christensen}}, \bibinfo
  {author} {\bibfnamefont {J.}~\bibnamefont {Larsen}}, \bibinfo {author}
  {\bibfnamefont {J.}~\bibnamefont {Mesot}}, \bibinfo {author} {\bibfnamefont
  {R.}~\bibnamefont {Liang}}, \bibinfo {author} {\bibfnamefont {D.~A.}\
  \bibnamefont {Bonn}}, \bibinfo {author} {\bibfnamefont {W.~N.}\ \bibnamefont
  {Hardy}}, \bibinfo {author} {\bibfnamefont {A.}~\bibnamefont {Watenphul}},
  \bibinfo {author} {\bibfnamefont {M.~v.}\ \bibnamefont {Zimmermann}},
  \bibinfo {author} {\bibfnamefont {E.~M.}\ \bibnamefont {Forgan}}, \ and\
  \bibinfo {author} {\bibfnamefont {S.~M.}\ \bibnamefont {Hayden}},\ }\href
  {\doibase 10.1038/nphys2456} {\bibfield  {journal} {\bibinfo  {journal}
  {Nature Physics}\ }\textbf {\bibinfo {volume} {8}},\ \bibinfo {pages} {871}
  (\bibinfo {year} {2012})},\ \Eprint {http://arxiv.org/abs/1206.4333}
  {arXiv:1206.4333} \BibitemShut {NoStop}%
\bibitem [{\citenamefont {Tabis}\ \emph {et~al.}(2014)\citenamefont {Tabis},
  \citenamefont {Li}, \citenamefont {{Le Tacon}}, \citenamefont {Braicovich},
  \citenamefont {Kreyssig}, \citenamefont {Minola}, \citenamefont {Dellea},
  \citenamefont {Weschke}, \citenamefont {Veit}, \citenamefont {Ramazanoglu},
  \citenamefont {Goldman}, \citenamefont {Schmitt}, \citenamefont
  {Ghiringhelli}, \citenamefont {Barisic}, \citenamefont {Chan}, \citenamefont
  {Dorow}, \citenamefont {Yu}, \citenamefont {Zhao}, \citenamefont {Keimer},\
  and\ \citenamefont {Greven}}]{Tabis2014}%
  \BibitemOpen
  \bibfield  {author} {\bibinfo {author} {\bibfnamefont {W.}~\bibnamefont
  {Tabis}}, \bibinfo {author} {\bibfnamefont {Y.}~\bibnamefont {Li}}, \bibinfo
  {author} {\bibfnamefont {M.}~\bibnamefont {{Le Tacon}}}, \bibinfo {author}
  {\bibfnamefont {L.}~\bibnamefont {Braicovich}}, \bibinfo {author}
  {\bibfnamefont {A.}~\bibnamefont {Kreyssig}}, \bibinfo {author}
  {\bibfnamefont {M.}~\bibnamefont {Minola}}, \bibinfo {author} {\bibfnamefont
  {G.}~\bibnamefont {Dellea}}, \bibinfo {author} {\bibfnamefont
  {E.}~\bibnamefont {Weschke}}, \bibinfo {author} {\bibfnamefont {M.~J.}\
  \bibnamefont {Veit}}, \bibinfo {author} {\bibfnamefont {M.}~\bibnamefont
  {Ramazanoglu}}, \bibinfo {author} {\bibfnamefont {A.~I.}\ \bibnamefont
  {Goldman}}, \bibinfo {author} {\bibfnamefont {T.}~\bibnamefont {Schmitt}},
  \bibinfo {author} {\bibfnamefont {G.}~\bibnamefont {Ghiringhelli}}, \bibinfo
  {author} {\bibfnamefont {N.}~\bibnamefont {Barisic}}, \bibinfo {author}
  {\bibfnamefont {M.~K.}\ \bibnamefont {Chan}}, \bibinfo {author}
  {\bibfnamefont {C.~J.}\ \bibnamefont {Dorow}}, \bibinfo {author}
  {\bibfnamefont {G.}~\bibnamefont {Yu}}, \bibinfo {author} {\bibfnamefont
  {X.}~\bibnamefont {Zhao}}, \bibinfo {author} {\bibfnamefont {B.}~\bibnamefont
  {Keimer}}, \ and\ \bibinfo {author} {\bibfnamefont {M.}~\bibnamefont
  {Greven}},\ }\href {\doibase 10.1038/ncomms6875} {\bibfield  {journal}
  {\bibinfo  {journal} {Nature Communications}\ }\textbf {\bibinfo {volume}
  {5}},\ \bibinfo {pages} {5875} (\bibinfo {year} {2014})}\BibitemShut
  {NoStop}%
\bibitem [{\citenamefont {Sebastian}\ \emph {et~al.}(2014)\citenamefont
  {Sebastian}, \citenamefont {Harrison}, \citenamefont {Balakirev},
  \citenamefont {Altarawneh}, \citenamefont {Goddard}, \citenamefont {Liang},
  \citenamefont {Bonn}, \citenamefont {Hardy},\ and\ \citenamefont
  {Lonzarich}}]{Sebastian2014}%
  \BibitemOpen
  \bibfield  {author} {\bibinfo {author} {\bibfnamefont {S.~E.}\ \bibnamefont
  {Sebastian}}, \bibinfo {author} {\bibfnamefont {N.}~\bibnamefont {Harrison}},
  \bibinfo {author} {\bibfnamefont {F.~F.}\ \bibnamefont {Balakirev}}, \bibinfo
  {author} {\bibfnamefont {M.~M.}\ \bibnamefont {Altarawneh}}, \bibinfo
  {author} {\bibfnamefont {P.~A.}\ \bibnamefont {Goddard}}, \bibinfo {author}
  {\bibfnamefont {R.}~\bibnamefont {Liang}}, \bibinfo {author} {\bibfnamefont
  {D.~A.}\ \bibnamefont {Bonn}}, \bibinfo {author} {\bibfnamefont {W.~N.}\
  \bibnamefont {Hardy}}, \ and\ \bibinfo {author} {\bibfnamefont {G.~G.}\
  \bibnamefont {Lonzarich}},\ }\href {\doibase 10.1038/nature13326} {\bibfield
  {journal} {\bibinfo  {journal} {Nature}\ }\textbf {\bibinfo {volume} {511}},\
  \bibinfo {pages} {61} (\bibinfo {year} {2014})}\BibitemShut {NoStop}%
\bibitem [{\citenamefont {Gerber}\ \emph {et~al.}(2015)\citenamefont {Gerber},
  \citenamefont {Jang}, \citenamefont {Nojiri}, \citenamefont {Matsuzawa},
  \citenamefont {Yasamura}, \citenamefont {Bonn}, \citenamefont {Liang},
  \citenamefont {Hardy}, \citenamefont {Islam}, \citenamefont {Mehta},
  \citenamefont {Song}, \citenamefont {Sikorski}, \citenamefont {Stefanescu},
  \citenamefont {Feng}, \citenamefont {Kivelson}, \citenamefont {Devereaux},
  \citenamefont {Shen}, \citenamefont {Kao}, \citenamefont {Lee}, \citenamefont
  {Zhu},\ and\ \citenamefont {Lee}}]{Gerber2015}%
  \BibitemOpen
  \bibfield  {author} {\bibinfo {author} {\bibfnamefont {S.}~\bibnamefont
  {Gerber}}, \bibinfo {author} {\bibfnamefont {H.}~\bibnamefont {Jang}},
  \bibinfo {author} {\bibfnamefont {H.}~\bibnamefont {Nojiri}}, \bibinfo
  {author} {\bibfnamefont {S.}~\bibnamefont {Matsuzawa}}, \bibinfo {author}
  {\bibfnamefont {H.}~\bibnamefont {Yasamura}}, \bibinfo {author}
  {\bibfnamefont {D.~A.}\ \bibnamefont {Bonn}}, \bibinfo {author}
  {\bibfnamefont {R.}~\bibnamefont {Liang}}, \bibinfo {author} {\bibfnamefont
  {W.~N.}\ \bibnamefont {Hardy}}, \bibinfo {author} {\bibfnamefont
  {Z.}~\bibnamefont {Islam}}, \bibinfo {author} {\bibfnamefont
  {A.}~\bibnamefont {Mehta}}, \bibinfo {author} {\bibfnamefont
  {S.}~\bibnamefont {Song}}, \bibinfo {author} {\bibfnamefont {M.}~\bibnamefont
  {Sikorski}}, \bibinfo {author} {\bibfnamefont {D.}~\bibnamefont
  {Stefanescu}}, \bibinfo {author} {\bibfnamefont {Y.}~\bibnamefont {Feng}},
  \bibinfo {author} {\bibfnamefont {S.~A.}\ \bibnamefont {Kivelson}}, \bibinfo
  {author} {\bibfnamefont {T.~P.}\ \bibnamefont {Devereaux}}, \bibinfo {author}
  {\bibfnamefont {Z.-X.}\ \bibnamefont {Shen}}, \bibinfo {author}
  {\bibfnamefont {C.-C.}\ \bibnamefont {Kao}}, \bibinfo {author} {\bibfnamefont
  {W.-S.}\ \bibnamefont {Lee}}, \bibinfo {author} {\bibfnamefont
  {D.}~\bibnamefont {Zhu}}, \ and\ \bibinfo {author} {\bibfnamefont {J.-S.}\
  \bibnamefont {Lee}},\ }\href@noop {} {\bibfield  {journal} {\bibinfo
  {journal} {Science}\ }\textbf {\bibinfo {volume} {350}},\ \bibinfo {pages}
  {949} (\bibinfo {year} {2015})}\BibitemShut {NoStop}%
\bibitem [{\citenamefont {Maharaj}\ \emph {et~al.}(2016)\citenamefont
  {Maharaj}, \citenamefont {Zhang}, \citenamefont {Ramshaw},\ and\
  \citenamefont {Kivelson}}]{Maharaj2016}%
  \BibitemOpen
  \bibfield  {author} {\bibinfo {author} {\bibfnamefont {A.~V.}\ \bibnamefont
  {Maharaj}}, \bibinfo {author} {\bibfnamefont {Y.}~\bibnamefont {Zhang}},
  \bibinfo {author} {\bibfnamefont {B.~J.}\ \bibnamefont {Ramshaw}}, \ and\
  \bibinfo {author} {\bibfnamefont {S.~A.}\ \bibnamefont {Kivelson}},\ }\href
  {\doibase 10.1103/PhysRevB.93.094503} {\bibfield  {journal} {\bibinfo
  {journal} {Physical Review B}\ }\textbf {\bibinfo {volume} {93}},\ \bibinfo
  {pages} {094503} (\bibinfo {year} {2016})}\BibitemShut {NoStop}%
\bibitem [{\citenamefont {Briffa}\ \emph {et~al.}(2016)\citenamefont {Briffa},
  \citenamefont {Blackburn}, \citenamefont {Hayden}, \citenamefont {Yelland},
  \citenamefont {Long},\ and\ \citenamefont {Forgan}}]{Briffa2016}%
  \BibitemOpen
  \bibfield  {author} {\bibinfo {author} {\bibfnamefont {A.~K.~R.}\
  \bibnamefont {Briffa}}, \bibinfo {author} {\bibfnamefont {E.}~\bibnamefont
  {Blackburn}}, \bibinfo {author} {\bibfnamefont {S.~M.}\ \bibnamefont
  {Hayden}}, \bibinfo {author} {\bibfnamefont {E.~A.}\ \bibnamefont {Yelland}},
  \bibinfo {author} {\bibfnamefont {M.~W.}\ \bibnamefont {Long}}, \ and\
  \bibinfo {author} {\bibfnamefont {E.~M.}\ \bibnamefont {Forgan}},\ }\href
  {\doibase 10.1103/PhysRevB.93.094502} {\bibfield  {journal} {\bibinfo
  {journal} {Physical Review B}\ }\textbf {\bibinfo {volume} {93}},\ \bibinfo
  {pages} {094502} (\bibinfo {year} {2016})}\BibitemShut {NoStop}%
\bibitem [{\citenamefont {Forgan}\ \emph {et~al.}(2015)\citenamefont {Forgan},
  \citenamefont {Blackburn}, \citenamefont {Holmes}, \citenamefont {Briffa},
  \citenamefont {Chang}, \citenamefont {Bouchenoire}, \citenamefont {Brown},
  \citenamefont {Liang}, \citenamefont {Bonn}, \citenamefont {Hardy},
  \citenamefont {Christensen}, \citenamefont {Zimmermann}, \citenamefont
  {H{\"{u}}cker},\ and\ \citenamefont {Hayden}}]{Forgan2015}%
  \BibitemOpen
  \bibfield  {author} {\bibinfo {author} {\bibfnamefont {E.~M.}\ \bibnamefont
  {Forgan}}, \bibinfo {author} {\bibfnamefont {E.}~\bibnamefont {Blackburn}},
  \bibinfo {author} {\bibfnamefont {A.~T.}\ \bibnamefont {Holmes}}, \bibinfo
  {author} {\bibfnamefont {A.~K.~R.}\ \bibnamefont {Briffa}}, \bibinfo {author}
  {\bibfnamefont {J.}~\bibnamefont {Chang}}, \bibinfo {author} {\bibfnamefont
  {L.}~\bibnamefont {Bouchenoire}}, \bibinfo {author} {\bibfnamefont {S.~D.}\
  \bibnamefont {Brown}}, \bibinfo {author} {\bibfnamefont {R.}~\bibnamefont
  {Liang}}, \bibinfo {author} {\bibfnamefont {D.}~\bibnamefont {Bonn}},
  \bibinfo {author} {\bibfnamefont {W.~N.}\ \bibnamefont {Hardy}}, \bibinfo
  {author} {\bibfnamefont {N.~B.}\ \bibnamefont {Christensen}}, \bibinfo
  {author} {\bibfnamefont {M.~v.}\ \bibnamefont {Zimmermann}}, \bibinfo
  {author} {\bibfnamefont {M.}~\bibnamefont {H{\"{u}}cker}}, \ and\ \bibinfo
  {author} {\bibfnamefont {S.~M.}\ \bibnamefont {Hayden}},\ }\href {\doibase
  10.1038/ncomms10064} {\bibfield  {journal} {\bibinfo  {journal} {Nature
  Communications}\ }\textbf {\bibinfo {volume} {6}},\ \bibinfo {pages} {10064}
  (\bibinfo {year} {2015})}\BibitemShut {NoStop}%
\bibitem [{\citenamefont {Kennett}\ and\ \citenamefont
  {McKenzie}(2007)}]{Kennett2007}%
  \BibitemOpen
  \bibfield  {author} {\bibinfo {author} {\bibfnamefont {M.~P.}\ \bibnamefont
  {Kennett}}\ and\ \bibinfo {author} {\bibfnamefont {R.~H.}\ \bibnamefont
  {McKenzie}},\ }\href {\doibase 10.1103/PhysRevB.76.054515} {\bibfield
  {journal} {\bibinfo  {journal} {Physical Review B}\ }\textbf {\bibinfo
  {volume} {76}},\ \bibinfo {pages} {054515} (\bibinfo {year} {2007})},\
  \Eprint {http://arxiv.org/abs/0610191} {arXiv:0610191 [cond-mat]}
  \BibitemShut {NoStop}%
\bibitem [{Note1()}]{Note1}%
  \BibitemOpen
  \bibinfo {note} {Coherent and weakly incoherent interlayer transport will
  have different signatures in ADMR only when the applied field is nearly
  in-plane, which is a regime that we have not included in these
  calculations\cite {Kennett2007}}\BibitemShut {NoStop}%
\bibitem [{\citenamefont {Bari{\v{s}}i{\'{c}}}\ \emph
  {et~al.}(2008)\citenamefont {Bari{\v{s}}i{\'{c}}}, \citenamefont {Li},
  \citenamefont {Zhao}, \citenamefont {Cho}, \citenamefont {Chabot-Couture},
  \citenamefont {Yu},\ and\ \citenamefont {Greven}}]{Barisic2008}%
  \BibitemOpen
  \bibfield  {author} {\bibinfo {author} {\bibfnamefont {N.}~\bibnamefont
  {Bari{\v{s}}i{\'{c}}}}, \bibinfo {author} {\bibfnamefont {Y.}~\bibnamefont
  {Li}}, \bibinfo {author} {\bibfnamefont {X.}~\bibnamefont {Zhao}}, \bibinfo
  {author} {\bibfnamefont {Y.-C.}\ \bibnamefont {Cho}}, \bibinfo {author}
  {\bibfnamefont {G.}~\bibnamefont {Chabot-Couture}}, \bibinfo {author}
  {\bibfnamefont {G.}~\bibnamefont {Yu}}, \ and\ \bibinfo {author}
  {\bibfnamefont {M.}~\bibnamefont {Greven}},\ }\href {\doibase
  10.1103/PhysRevB.78.054518} {\bibfield  {journal} {\bibinfo  {journal}
  {Physical Review B}\ }\textbf {\bibinfo {volume} {78}},\ \bibinfo {pages}
  {054518} (\bibinfo {year} {2008})},\ \Eprint {http://arxiv.org/abs/0805.1497}
  {arXiv:0805.1497} \BibitemShut {NoStop}%
\bibitem [{\citenamefont {Ramshaw}\ \emph {et~al.}(2017)\citenamefont
  {Ramshaw}, \citenamefont {Harrison}, \citenamefont {Sebastian}, \citenamefont
  {Ghannadzadeh}, \citenamefont {Modic}, \citenamefont {Bonn}, \citenamefont
  {Hardy}, \citenamefont {Liang},\ and\ \citenamefont {Goddard}}]{Ramshaw2017}%
  \BibitemOpen
  \bibfield  {author} {\bibinfo {author} {\bibfnamefont {B.~J.}\ \bibnamefont
  {Ramshaw}}, \bibinfo {author} {\bibfnamefont {N.}~\bibnamefont {Harrison}},
  \bibinfo {author} {\bibfnamefont {S.~E.}\ \bibnamefont {Sebastian}}, \bibinfo
  {author} {\bibfnamefont {S.}~\bibnamefont {Ghannadzadeh}}, \bibinfo {author}
  {\bibfnamefont {K.~A.}\ \bibnamefont {Modic}}, \bibinfo {author}
  {\bibfnamefont {D.~A.}\ \bibnamefont {Bonn}}, \bibinfo {author}
  {\bibfnamefont {W.~N.}\ \bibnamefont {Hardy}}, \bibinfo {author}
  {\bibfnamefont {R.}~\bibnamefont {Liang}}, \ and\ \bibinfo {author}
  {\bibfnamefont {P.~A.}\ \bibnamefont {Goddard}},\ }\href {\doibase
  10.1038/s41535-017-0013-z} {\bibfield  {journal} {\bibinfo  {journal}
  {Quantum Materials}\ }\textbf {\bibinfo {volume} {2}} (\bibinfo {year}
  {2017}),\ 10.1038/s41535-017-0013-z},\ \Eprint
  {http://arxiv.org/abs/1607.07145} {arXiv:1607.07145} \BibitemShut {NoStop}%
\bibitem [{\citenamefont {Ziman}(1972)}]{Ziman1972}%
  \BibitemOpen
  \bibfield  {author} {\bibinfo {author} {\bibfnamefont {J.~M.}\ \bibnamefont
  {Ziman}},\ }\href@noop {} {\emph {\bibinfo {title} {{Principles of the Theory
  of Solids}}}}\ (\bibinfo  {publisher} {Cambridge University Press},\ \bibinfo
  {year} {1972})\ p.\ \bibinfo {pages} {301}\BibitemShut {NoStop}%
\bibitem [{\citenamefont {Abdel-Jawad}\ \emph {et~al.}(2006)\citenamefont
  {Abdel-Jawad}, \citenamefont {Kennett}, \citenamefont {Balicas},
  \citenamefont {Carrington}, \citenamefont {Mackenzie}, \citenamefont
  {McKenzie},\ and\ \citenamefont {Hussey}}]{Abdel-Jawad2006}%
  \BibitemOpen
  \bibfield  {author} {\bibinfo {author} {\bibfnamefont {M.}~\bibnamefont
  {Abdel-Jawad}}, \bibinfo {author} {\bibfnamefont {M.~P.}\ \bibnamefont
  {Kennett}}, \bibinfo {author} {\bibfnamefont {L.}~\bibnamefont {Balicas}},
  \bibinfo {author} {\bibfnamefont {A.}~\bibnamefont {Carrington}}, \bibinfo
  {author} {\bibfnamefont {A.~P.}\ \bibnamefont {Mackenzie}}, \bibinfo {author}
  {\bibfnamefont {R.~H.}\ \bibnamefont {McKenzie}}, \ and\ \bibinfo {author}
  {\bibfnamefont {N.~E.}\ \bibnamefont {Hussey}},\ }\href {\doibase
  10.1038/nphys449} {\bibfield  {journal} {\bibinfo  {journal} {Nature
  Physics}\ }\textbf {\bibinfo {volume} {2}},\ \bibinfo {pages} {821} (\bibinfo
  {year} {2006})}\BibitemShut {NoStop}%
\bibitem [{Note2()}]{Note2}%
  \BibitemOpen
  \bibinfo {note} {This assumes that the Brillouin zone height is $\protect
  \frac {2\pi }{c}$. In, e.g., a body-centered tetragonal structure with
  Brillouin zone height $\protect \frac {4\pi }{c}$ the cosine term would be
  $\protect \qopname \relax o{cos}\left (\protect \frac {ck_z}{2}\right
  )$.}\BibitemShut {Stop}%
\bibitem [{\citenamefont {Cohen}\ and\ \citenamefont
  {Falicov}(1961)}]{Cohen1961}%
  \BibitemOpen
  \bibfield  {author} {\bibinfo {author} {\bibfnamefont {M.~H.}\ \bibnamefont
  {Cohen}}\ and\ \bibinfo {author} {\bibfnamefont {L.~M.}\ \bibnamefont
  {Falicov}},\ }\href@noop {} {\bibfield  {journal} {\bibinfo  {journal}
  {Physical Review Letters}\ }\textbf {\bibinfo {volume} {7}},\ \bibinfo
  {pages} {231} (\bibinfo {year} {1961})}\BibitemShut {NoStop}%
\bibitem [{\citenamefont {Shoenberg}(1984)}]{Shoenberg}%
  \BibitemOpen
  \bibfield  {author} {\bibinfo {author} {\bibfnamefont {D.}~\bibnamefont
  {Shoenberg}},\ }\href
  {http://ebooks.cambridge.org/ebook.jsf?bid=CBO9780511897870} {\emph {\bibinfo
  {title} {{Magnetic Oscillations in Metals}}}}\ (\bibinfo  {publisher}
  {Cambridge University Press},\ \bibinfo {year} {1984})\ pp.\ \bibinfo {pages}
  {331--368}\BibitemShut {NoStop}%
\bibitem [{\citenamefont {Lewin}\ and\ \citenamefont
  {Analytis}(2015)}]{Lewin2015a}%
  \BibitemOpen
  \bibfield  {author} {\bibinfo {author} {\bibfnamefont {S.~K.}\ \bibnamefont
  {Lewin}}\ and\ \bibinfo {author} {\bibfnamefont {J.~G.}\ \bibnamefont
  {Analytis}},\ }\href {\doibase 10.1103/PhysRevB.92.195130} {\bibfield
  {journal} {\bibinfo  {journal} {Physical Review B}\ }\textbf {\bibinfo
  {volume} {92}},\ \bibinfo {pages} {195130} (\bibinfo {year} {2015})},\
  \Eprint {http://arxiv.org/abs/1507.08635} {arXiv:1507.08635} \BibitemShut
  {NoStop}%
\bibitem [{\citenamefont {Harrison}\ and\ \citenamefont
  {Sebastian}(2014)}]{Harrison2014}%
  \BibitemOpen
  \bibfield  {author} {\bibinfo {author} {\bibfnamefont {N.}~\bibnamefont
  {Harrison}}\ and\ \bibinfo {author} {\bibfnamefont {S.~E.}\ \bibnamefont
  {Sebastian}},\ }\href {\doibase 10.1088/1367-2630/16/6/063025} {\bibfield
  {journal} {\bibinfo  {journal} {New Journal of Physics}\ }\textbf {\bibinfo
  {volume} {16}},\ \bibinfo {pages} {063025} (\bibinfo {year} {2014})},\
  \Eprint {http://arxiv.org/abs/1401.6590} {arXiv:1401.6590} \BibitemShut
  {NoStop}%
\bibitem [{\citenamefont {Falicov}\ and\ \citenamefont
  {Sievert}(1965)}]{Falicov1965}%
  \BibitemOpen
  \bibfield  {author} {\bibinfo {author} {\bibfnamefont {L.}~\bibnamefont
  {Falicov}}\ and\ \bibinfo {author} {\bibfnamefont {P.}~\bibnamefont
  {Sievert}},\ }\href {\doibase 10.1103/PhysRev.95.31} {\bibfield  {journal}
  {\bibinfo  {journal} {Physical Review}\ }\textbf {\bibinfo {volume} {138}},\
  \bibinfo {pages} {A88} (\bibinfo {year} {1965})}\BibitemShut {NoStop}%
\bibitem [{\citenamefont {Nowojewski}\ \emph {et~al.}(2008)\citenamefont
  {Nowojewski}, \citenamefont {Goddard},\ and\ \citenamefont
  {Blundell}}]{Nowojewski2008}%
  \BibitemOpen
  \bibfield  {author} {\bibinfo {author} {\bibfnamefont {A.}~\bibnamefont
  {Nowojewski}}, \bibinfo {author} {\bibfnamefont {P.}~\bibnamefont {Goddard}},
  \ and\ \bibinfo {author} {\bibfnamefont {S.~J.}\ \bibnamefont {Blundell}},\
  }\href {\doibase 10.1103/PhysRevB.77.012402} {\bibfield  {journal} {\bibinfo
  {journal} {Physical Review B}\ }\textbf {\bibinfo {volume} {77}},\ \bibinfo
  {pages} {012402} (\bibinfo {year} {2008})}\BibitemShut {NoStop}%
\bibitem [{\citenamefont {Nowojewski}\ and\ \citenamefont
  {Blundell}(2010)}]{Nowojewski2010}%
  \BibitemOpen
  \bibfield  {author} {\bibinfo {author} {\bibfnamefont {A.}~\bibnamefont
  {Nowojewski}}\ and\ \bibinfo {author} {\bibfnamefont {S.~J.}\ \bibnamefont
  {Blundell}},\ }\href {\doibase 10.1103/PhysRevB.82.075121} {\bibfield
  {journal} {\bibinfo  {journal} {Physical Review B}\ }\textbf {\bibinfo
  {volume} {82}},\ \bibinfo {pages} {075121} (\bibinfo {year}
  {2010})}\BibitemShut {NoStop}%
\bibitem [{\citenamefont {Blundell}\ \emph {et~al.}(2010)\citenamefont
  {Blundell}, \citenamefont {Nowojewski},\ and\ \citenamefont
  {Goddard}}]{Blundell2010}%
  \BibitemOpen
  \bibfield  {author} {\bibinfo {author} {\bibfnamefont {S.~J.}\ \bibnamefont
  {Blundell}}, \bibinfo {author} {\bibfnamefont {A.}~\bibnamefont
  {Nowojewski}}, \ and\ \bibinfo {author} {\bibfnamefont {P.~A.}\ \bibnamefont
  {Goddard}},\ }\href {\doibase 10.1016/j.physb.2009.11.008} {\bibfield
  {journal} {\bibinfo  {journal} {Physica B: Condensed Matter}\ }\textbf
  {\bibinfo {volume} {405}},\ \bibinfo {pages} {S134} (\bibinfo {year}
  {2010})}\BibitemShut {NoStop}%
\bibitem [{\citenamefont {Zhao}\ \emph {et~al.}(2006)\citenamefont {Zhao},
  \citenamefont {Yu}, \citenamefont {Cho}, \citenamefont {Chabot-Couture},
  \citenamefont {Bari{\v{s}}ic}, \citenamefont {Bourges}, \citenamefont
  {Kaneko}, \citenamefont {Li}, \citenamefont {Lu}, \citenamefont {Motoyama},
  \citenamefont {Vajk},\ and\ \citenamefont {Greven}}]{Zhao2006}%
  \BibitemOpen
  \bibfield  {author} {\bibinfo {author} {\bibfnamefont {X.}~\bibnamefont
  {Zhao}}, \bibinfo {author} {\bibfnamefont {G.}~\bibnamefont {Yu}}, \bibinfo
  {author} {\bibfnamefont {Y.-c.}\ \bibnamefont {Cho}}, \bibinfo {author}
  {\bibfnamefont {G.}~\bibnamefont {Chabot-Couture}}, \bibinfo {author}
  {\bibfnamefont {N.}~\bibnamefont {Bari{\v{s}}ic}}, \bibinfo {author}
  {\bibfnamefont {P.}~\bibnamefont {Bourges}}, \bibinfo {author} {\bibfnamefont
  {N.}~\bibnamefont {Kaneko}}, \bibinfo {author} {\bibfnamefont
  {Y.}~\bibnamefont {Li}}, \bibinfo {author} {\bibfnamefont {L.}~\bibnamefont
  {Lu}}, \bibinfo {author} {\bibfnamefont {E.~M.}\ \bibnamefont {Motoyama}},
  \bibinfo {author} {\bibfnamefont {O.~P.}\ \bibnamefont {Vajk}}, \ and\
  \bibinfo {author} {\bibfnamefont {M.}~\bibnamefont {Greven}},\ }\href@noop {}
  {\bibfield  {journal} {\bibinfo  {journal} {Advanced Material}\ }\textbf
  {\bibinfo {volume} {18}},\ \bibinfo {pages} {3243} (\bibinfo {year}
  {2006})}\BibitemShut {NoStop}%
\bibitem [{Note3()}]{Note3}%
  \BibitemOpen
  \bibinfo {note} {The constraint used for the cross-sectional Fermi surface
  area is from a sample that also had an as-grown $T_c$ of roughly 80 K but
  that was subsequently heat-treated in a nitrogen-rich atmosphere to achieve a
  $T_c$ of 74 K~\cite {Chan2016}. This treatment may slightly decrease the
  interlayer lattice parameter since it removes interstitial oxygen, but such
  an effect should be neglible: only a fraction of a percent change, based on
  similar behavior in YBCO~\cite {Liang2006}.}\BibitemShut {Stop}%
\bibitem [{\citenamefont {Prentice}\ and\ \citenamefont
  {Coldea}(2016)}]{Prentice2016}%
  \BibitemOpen
  \bibfield  {author} {\bibinfo {author} {\bibfnamefont {J.~C.~A.}\
  \bibnamefont {Prentice}}\ and\ \bibinfo {author} {\bibfnamefont {A.~I.}\
  \bibnamefont {Coldea}},\ }\href {\doibase 10.1103/PhysRevB.93.245105}
  {\bibfield  {journal} {\bibinfo  {journal} {Physical Review B}\ }\textbf
  {\bibinfo {volume} {93}},\ \bibinfo {pages} {245105} (\bibinfo {year}
  {2016})}\BibitemShut {NoStop}%
\bibitem [{\citenamefont {Hussey}\ \emph {et~al.}(2003)\citenamefont {Hussey},
  \citenamefont {Abdel-Jawad}, \citenamefont {Carrington}, \citenamefont
  {Mackenzie},\ and\ \citenamefont {Balicas}}]{Hussey2003}%
  \BibitemOpen
  \bibfield  {author} {\bibinfo {author} {\bibfnamefont {N.~E.}\ \bibnamefont
  {Hussey}}, \bibinfo {author} {\bibfnamefont {M.}~\bibnamefont {Abdel-Jawad}},
  \bibinfo {author} {\bibfnamefont {A.}~\bibnamefont {Carrington}}, \bibinfo
  {author} {\bibfnamefont {A.~P.}\ \bibnamefont {Mackenzie}}, \ and\ \bibinfo
  {author} {\bibfnamefont {L.}~\bibnamefont {Balicas}},\ }\href {\doibase
  10.1038/nature02071.1.} {\bibfield  {journal} {\bibinfo  {journal} {Nature}\
  }\textbf {\bibinfo {volume} {425}},\ \bibinfo {pages} {814} (\bibinfo {year}
  {2003})}\BibitemShut {NoStop}%
\bibitem [{\citenamefont {Bergemann}\ \emph {et~al.}(2003)\citenamefont
  {Bergemann}, \citenamefont {Mackenzie}, \citenamefont {Julian}, \citenamefont
  {Forsythe},\ and\ \citenamefont {Ohmichi}}]{Bergemann2003}%
  \BibitemOpen
  \bibfield  {author} {\bibinfo {author} {\bibfnamefont {C.}~\bibnamefont
  {Bergemann}}, \bibinfo {author} {\bibfnamefont {A.~P.}\ \bibnamefont
  {Mackenzie}}, \bibinfo {author} {\bibfnamefont {S.~R.}\ \bibnamefont
  {Julian}}, \bibinfo {author} {\bibfnamefont {D.}~\bibnamefont {Forsythe}}, \
  and\ \bibinfo {author} {\bibfnamefont {E.}~\bibnamefont {Ohmichi}},\ }\href
  {\doibase 10.1080/00018730310001621737} {\bibfield  {journal} {\bibinfo
  {journal} {Advances in Physics}\ }\textbf {\bibinfo {volume} {52}},\ \bibinfo
  {pages} {639} (\bibinfo {year} {2003})}\BibitemShut {NoStop}%
\bibitem [{Note4()}]{Note4}%
  \BibitemOpen
  \bibinfo {note} {This is the average result across samples with $T_c = 71 K$
  and $T_c = 74 K$.}\BibitemShut {Stop}%
\bibitem [{\citenamefont {Grissonnanche}\ \emph {et~al.}(2014)\citenamefont
  {Grissonnanche}, \citenamefont {Cyr-Choini{\`{e}}re}, \citenamefont
  {Lalibert{\'{e}}}, \citenamefont {{Ren{\'{e}} de Cotret}}, \citenamefont
  {Juneau-Fecteau}, \citenamefont {Dufour-Beaus{\'{e}}jour}, \citenamefont
  {Delage}, \citenamefont {LeBoeuf}, \citenamefont {Chang}, \citenamefont
  {Ramshaw}, \citenamefont {Bonn}, \citenamefont {Hardy}, \citenamefont
  {Liang}, \citenamefont {Adachi}, \citenamefont {Hussey}, \citenamefont
  {Vignolle}, \citenamefont {Proust}, \citenamefont {Sutherland}, \citenamefont
  {Kr{\"{a}}mer}, \citenamefont {Park}, \citenamefont {Graf}, \citenamefont
  {Doiron-Leyraud},\ and\ \citenamefont {Taillefer}}]{Grissonnanche2014}%
  \BibitemOpen
  \bibfield  {author} {\bibinfo {author} {\bibfnamefont {G.}~\bibnamefont
  {Grissonnanche}}, \bibinfo {author} {\bibfnamefont {O.}~\bibnamefont
  {Cyr-Choini{\`{e}}re}}, \bibinfo {author} {\bibfnamefont {F.}~\bibnamefont
  {Lalibert{\'{e}}}}, \bibinfo {author} {\bibfnamefont {S.}~\bibnamefont
  {{Ren{\'{e}} de Cotret}}}, \bibinfo {author} {\bibfnamefont {A.}~\bibnamefont
  {Juneau-Fecteau}}, \bibinfo {author} {\bibfnamefont {S.}~\bibnamefont
  {Dufour-Beaus{\'{e}}jour}}, \bibinfo {author} {\bibfnamefont {M.-{\`{E}}.}\
  \bibnamefont {Delage}}, \bibinfo {author} {\bibfnamefont {D.}~\bibnamefont
  {LeBoeuf}}, \bibinfo {author} {\bibfnamefont {J.}~\bibnamefont {Chang}},
  \bibinfo {author} {\bibfnamefont {B.~J.}\ \bibnamefont {Ramshaw}}, \bibinfo
  {author} {\bibfnamefont {D.~A.}\ \bibnamefont {Bonn}}, \bibinfo {author}
  {\bibfnamefont {W.~N.}\ \bibnamefont {Hardy}}, \bibinfo {author}
  {\bibfnamefont {R.}~\bibnamefont {Liang}}, \bibinfo {author} {\bibfnamefont
  {S.}~\bibnamefont {Adachi}}, \bibinfo {author} {\bibfnamefont {N.~E.}\
  \bibnamefont {Hussey}}, \bibinfo {author} {\bibfnamefont {B.}~\bibnamefont
  {Vignolle}}, \bibinfo {author} {\bibfnamefont {C.}~\bibnamefont {Proust}},
  \bibinfo {author} {\bibfnamefont {M.}~\bibnamefont {Sutherland}}, \bibinfo
  {author} {\bibfnamefont {S.}~\bibnamefont {Kr{\"{a}}mer}}, \bibinfo {author}
  {\bibfnamefont {J.-H.}\ \bibnamefont {Park}}, \bibinfo {author}
  {\bibfnamefont {D.}~\bibnamefont {Graf}}, \bibinfo {author} {\bibfnamefont
  {N.}~\bibnamefont {Doiron-Leyraud}}, \ and\ \bibinfo {author} {\bibfnamefont
  {L.}~\bibnamefont {Taillefer}},\ }\href {\doibase 10.1038/ncomms4280}
  {\bibfield  {journal} {\bibinfo  {journal} {Nature Communications}\ }\textbf
  {\bibinfo {volume} {5}},\ \bibinfo {pages} {3280} (\bibinfo {year}
  {2014})}\BibitemShut {NoStop}%
\bibitem [{\citenamefont {Bulaevskii}(1990)}]{Bulaevskii1990}%
  \BibitemOpen
  \bibfield  {author} {\bibinfo {author} {\bibfnamefont {L.~N.}\ \bibnamefont
  {Bulaevskii}},\ }\href@noop {} {\bibfield  {journal} {\bibinfo  {journal}
  {International Journal of Modern Physics B}\ }\textbf {\bibinfo {volume}
  {4}},\ \bibinfo {pages} {1849} (\bibinfo {year} {1990})}\BibitemShut
  {NoStop}%
\bibitem [{\citenamefont {Schneider}\ and\ \citenamefont
  {Schmidt}(1993)}]{Schneider1993}%
  \BibitemOpen
  \bibfield  {author} {\bibinfo {author} {\bibfnamefont {T.}~\bibnamefont
  {Schneider}}\ and\ \bibinfo {author} {\bibfnamefont {A.}~\bibnamefont
  {Schmidt}},\ }\href {\doibase 10.1103/PhysRevB.47.5915} {\bibfield  {journal}
  {\bibinfo  {journal} {Physical Review B}\ }\textbf {\bibinfo {volume} {47}},\
  \bibinfo {pages} {5915} (\bibinfo {year} {1993})}\BibitemShut {NoStop}%
\bibitem [{\citenamefont {Tinkham}(1996)}]{Tinkham1996}%
  \BibitemOpen
  \bibfield  {author} {\bibinfo {author} {\bibfnamefont {M.}~\bibnamefont
  {Tinkham}},\ }\href@noop {} {\emph {\bibinfo {title} {{Introduction to
  Superconductivity}}}},\ \bibinfo {edition} {2nd}\ ed.\ (\bibinfo  {publisher}
  {McGraw-Hill, Inc.},\ \bibinfo {address} {New York City},\ \bibinfo {year}
  {1996})\ p.\ \bibinfo {pages} {139}\BibitemShut {NoStop}%
\bibitem [{\citenamefont {Lebed}\ \emph {et~al.}(2004)\citenamefont {Lebed},
  \citenamefont {Bagmet},\ and\ \citenamefont {Naughton}}]{Lebed2004}%
  \BibitemOpen
  \bibfield  {author} {\bibinfo {author} {\bibfnamefont {A.~G.}\ \bibnamefont
  {Lebed}}, \bibinfo {author} {\bibfnamefont {N.~N.}\ \bibnamefont {Bagmet}}, \
  and\ \bibinfo {author} {\bibfnamefont {M.~J.}\ \bibnamefont {Naughton}},\
  }\href@noop {} {\bibfield  {journal} {\bibinfo  {journal} {J. Phys. IV
  France}\ }\textbf {\bibinfo {volume} {114}},\ \bibinfo {pages} {77} (\bibinfo
  {year} {2004})}\BibitemShut {NoStop}%
\bibitem [{\citenamefont {Harrison}\ \emph {et~al.}(1996)\citenamefont
  {Harrison}, \citenamefont {Bogaerts}, \citenamefont {Reinders}, \citenamefont
  {Singleton}, \citenamefont {Blundell},\ and\ \citenamefont
  {Herlach}}]{Harrison1996}%
  \BibitemOpen
  \bibfield  {author} {\bibinfo {author} {\bibfnamefont {N.}~\bibnamefont
  {Harrison}}, \bibinfo {author} {\bibfnamefont {R.}~\bibnamefont {Bogaerts}},
  \bibinfo {author} {\bibfnamefont {P.~H.~P.}\ \bibnamefont {Reinders}},
  \bibinfo {author} {\bibfnamefont {J.}~\bibnamefont {Singleton}}, \bibinfo
  {author} {\bibfnamefont {S.~J.}\ \bibnamefont {Blundell}}, \ and\ \bibinfo
  {author} {\bibfnamefont {F.}~\bibnamefont {Herlach}},\ }\href@noop {}
  {\bibfield  {journal} {\bibinfo  {journal} {Physical Review B}\ }\textbf
  {\bibinfo {volume} {54}},\ \bibinfo {pages} {9977} (\bibinfo {year}
  {1996})}\BibitemShut {NoStop}%
\bibitem [{\citenamefont {Hussey}(2003)}]{Hussey2003a}%
  \BibitemOpen
  \bibfield  {author} {\bibinfo {author} {\bibfnamefont {N.~E.}\ \bibnamefont
  {Hussey}},\ }\href {\doibase 10.1140/epjb/e2003-00059-9} {\bibfield
  {journal} {\bibinfo  {journal} {The European Physical Journal B}\ }\textbf
  {\bibinfo {volume} {31}},\ \bibinfo {pages} {495} (\bibinfo {year}
  {2003})}\BibitemShut {NoStop}%
\bibitem [{\citenamefont {Yagi}(1992)}]{Yagi1992}%
  \BibitemOpen
  \bibfield  {author} {\bibinfo {author} {\bibfnamefont {R.}~\bibnamefont
  {Yagi}},\ }\emph {\bibinfo {title} {{Oscillatory Magnetoresistance in
  Modulated Structures}}},\ \href@noop {} {Ph.D. thesis},\ \bibinfo  {school}
  {University of Tokyo} (\bibinfo {year} {1992})\BibitemShut {NoStop}%
\bibitem [{\citenamefont {Liang}\ \emph {et~al.}(2006)\citenamefont {Liang},
  \citenamefont {Bonn},\ and\ \citenamefont {Hardy}}]{Liang2006}%
  \BibitemOpen
  \bibfield  {author} {\bibinfo {author} {\bibfnamefont {R.}~\bibnamefont
  {Liang}}, \bibinfo {author} {\bibfnamefont {D.~A.}\ \bibnamefont {Bonn}}, \
  and\ \bibinfo {author} {\bibfnamefont {W.~N.}\ \bibnamefont {Hardy}},\ }\href
  {\doibase 10.1103/PhysRevB.73.180505} {\bibfield  {journal} {\bibinfo
  {journal} {Physical Review B}\ }\textbf {\bibinfo {volume} {73}},\ \bibinfo
  {pages} {180505(R)} (\bibinfo {year} {2006})},\ \Eprint
  {http://arxiv.org/abs/0510674} {arXiv:0510674 [cond-mat]} \BibitemShut
  {NoStop}%
\end{thebibliography}
\end{document}